\documentclass[reprint, superscriptaddress, hidelinks, amsmath, amssymb, aps, floatfix, longbibliography]{revtex4-2}
\usepackage{graphicx}
\usepackage{dcolumn}
\usepackage{bm}
\usepackage{amsmath, physics}
\usepackage{dsfont}
\usepackage[mathlines]{lineno}
\usepackage{float}
\usepackage[caption=false, justification=justified]{subfig}
\usepackage{tikz}
\usetikzlibrary{shapes.geometric}
\usetikzlibrary{quantikz2}
\usepackage{rotating}
\usepackage{hyperref}
\usepackage{booktabs}

\begin{document}

\preprint{APS/123-QED}

\title{Exponential-SWAP Gate between Bosonic Motional Modes of Trapped Ions}

\author{Eugene Koh Jun Wei}
\thanks{These authors contributed equally}
\affiliation{Centre for Quantum Technologies,
National University of Singapore, 3 Science Drive 2, 117543 Singapore,
Singapore}

\author{Rongjie Zhang}
\thanks{These authors contributed equally}
\affiliation{Centre for Quantum Technologies,
National University of Singapore, 3 Science Drive 2, 117543 Singapore,
Singapore}

\author{Dzmitry Matsukevich}
\affiliation{Centre for Quantum Technologies,
National University of Singapore, 3 Science Drive 2, 117543 Singapore,
Singapore}
\affiliation{Department of Physics,
National University of Singapore, 2 Science Drive 3, 117551 Singapore,
Singapore}

\date{\today}

\begin{abstract}
The exponential-SWAP (eSWAP) gate generates a coherent superposition of identity and SWAP operations, with continuously tunable amplitudes. We experimentally implement this gate between two bosonic motional modes of trapped ions using spin-independent and spin-dependent beam splitter interactions. Measurements of selected slices of the joint characteristic function for distinct input states test the gate and demonstrate continuous control of the identity-SWAP superposition.  This implementation provides a tunable interaction for hybrid qubit-bosonic quantum information processing.
\end{abstract}

\maketitle

Experimental advances in trapped ions~\cite{de2022error,PhysRevLett.124.170502,nguyen_experimental_2021}, photonics~\cite{zhong2020quantum,madsen2022quantum,asavanant2019generation,konno2024logical,larsen2025integrated}, neutral atoms~\cite{shaw2025erasure,brown2023time,hartke2022quantum,lienhard2025generation}, and superconducting quantum circuits~\cite{gao_entanglement_2019,eickbusch2022fast,hu2019quantum,sivak2023real,brock2025quantum,heeres2015cavity,wang2016schrodinger,gao2018programmable} have established diverse platforms for bosonic quantum information processing. Their large Hilbert spaces provide computational resources for continuous-variable quantum computation (CVQC)~\cite{RevModPhys.77.513,PhysRevLett.82.1784}, while coupling them to discrete quantum systems enables hybrid architectures~\cite{kemper2025hybrid,HybridPRXQuantum,van2011optical}. These developments broaden the available approaches to quantum processor design and offer opportunities for efficient use of quantum hardware for hybrid quantum computation~\cite{sutherland2021universal,katz2023programmable,chen_quantum_2021,varona2024towards}. Together, they support applications in Gaussian boson sampling~\cite{zhong2020quantum,madsen2022quantum,liu2025robustquantumcomputationaladvantage,chen2023scalable}, bosonic quantum error correction~\cite{michael2016new,ofek2016extending,gottesman2000encoding,putterman2025hardware,campagne2020quantum,de2022error,fluhmann2019encoding}, quantum machine learning~\cite{nguyen2021quantum,jastrzebski2026adaptive}, quantum thermodynamics~\cite{maslennikov2019quantum,PhysRevLett.123.080602}, and spin-boson quantum simulation~\cite{zhang2018experimental}.

Trapped-ion systems offer a natural platform for these approaches with bosonic motional modes arising from their near-harmonic trapping potentials and Coulomb couplings. These modes can exhibit long coherence times \cite{jarlaud2021coherence, lucas2007long} and their quantum states can be coherently controlled through spin-motion interactions~\cite{RevModPhys.75.281, wineland1998experimental, haffner2008quantum}. The shared spin states can also mediate couplings between different motional modes~\cite{PhysRevLett.124.170502,katz2023programmable,jeon2025two}, which provide useful primitives for state transfer~\cite{PhysRevLett.124.170502,hou2024coherent}, entanglement generation~\cite{PhysRevLett.124.170502,hou2024coherent}, quantum state comparison~\cite{nguyen_experimental_2021}, quantum state engineering~\cite{saner2026generating,kienzler2015quantum}, and simulation of multimode dynamics~\cite{slim2024optomechanical,jeon2025two,jastrzebski2026adaptive,zhang2018experimental,chen2023scalable}. 

The exponential-SWAP (eSWAP) is one such operation, coherently superposing the identity operation, which leaves the states unchanged, and the SWAP operation, which exchanges the states of two modes. A single-qubit rotation on an auxiliary spin state continuously controls their relative amplitudes, enabling tunable entanglement between the modes. These features make eSWAP a useful primitive for multimode quantum information processing across bosonic platforms. Together with SWAP tests, eSWAP has been proposed as a route to universal quantum computation with arbitrary continuous-variable encodings ~\cite{lau_universal_2016}. It also serves as a building block for density-matrix exponentiation and quantum machine-learning tasks, including principal component analysis, matrix inversion, and vector-distance estimation~\cite{lloyd_quantum_2014,lau_quantum_2017}.

Although it has been experimentally realized between superconducting bosonic modes~\cite{gao_entanglement_2019}, its implementation with trapped ions has not been reported. In this Letter, we realize eSWAP by combining spin-independent beam splitter interactions, an exponential-parity operation constructed by spin-dependent beam splitter interactions, together with an auxiliary motional mode.

\begin{figure*}[!t]
\centering
\resizebox{\textwidth}{!}{%
   \begin{quantikz}[font = \huge]
    \lstick{$\ket{+}\;\,$}   &\gategroup[4,steps=6,style={dotted, rounded corners, inner ysep = 20},label style={label position=below, yshift=-0.9cm}]{eSWAP($\theta$)}& \ctrl[style = {inner sep = 3.5}]{1}&\qw&\gate[3]{eP\qty(\theta)} &\qw&\ctrl[style = {inner sep = 3.5}]{1}&\gate[1, style={inner ysep=12pt}]{R_y\left(- \frac{\pi}{2}\right)}&\gate[1, style={inner ysep=12pt}]{\mapsto\ket{g}}&\qw&\gategroup[4,steps=5,style={dotted, rounded corners, inner ysep = 20, inner xsep = 15pt},label style={label position=below, yshift=-0.9cm}]{Characteristic Function Measurement} &\gate[1, style={inner ysep=12pt}]{R_y\left(\phi\right)}&\ctrl[style = {diamond, inner sep = 3.5}]{1}&\ctrl[style = {diamond, inner sep = 3.5}]{2}&\meter[style = {scale = 2},label style={xshift = 1.9cm, yshift=-1.5cm}]{\expval{\hat{Z}}}\\
    \lstick{$\ket{\psi}_a$}&\qw & \gate[2]{XBS\qty(\frac{\pi}{2})} & \ghost{eP\qty(\theta)} &\qw&\qw &\gate[2]{XBS\qty(-\frac{\pi}{2})}&\qw&\qw&\qw&\qw&\qw&\gate[1, style={inner ysep=8pt}]{D\left(\frac{\alpha}{2}\right)}&\qw&\qw\\
    \lstick{$\ket{\varphi}_b$} &\qw & \ghost{BS(-\frac{\pi}{2})}  & \gate[2][3cm]{U_{BS}\left(-\pi,\Phi\right)} &\qw&\gate[2][3cm]{U_{BS}\left(\pi,\Phi\right)} & \ghost{BS(-\frac{\pi}{2})}&\ghost{eP\qty(\sigma_z\theta)} &\qw&\qw&\qw&\qw&\qw&\gate[1, style={inner ysep=4pt}]{D\left(\frac{\beta}{2}\right)}&\qw\\
    \lstick{$\ket{0}_c\,$} &\qw & \qw & \ghost{BS(-\pi)} &\qw & \ghost{BS(\pi)} &\qw&\qw&\qw&\qw&\qw&\qw&\qw&\qw&\qw
\end{quantikz} 
}
\caption{Circuit diagram for the eSWAP gate and joint characteristic-function
measurement. Modes $a$ and $b$ are the computational motional modes,
and mode $c$ is an auxiliary mode initialized in $|0\rangle_c$. The $R_y(-\frac{\pi}{2})$ rotation, followed by optical pumping, resets the spin to $\ket{g}$ before the measurement sequence. Circles denote operations conditioned on the spin in the $x$-basis and diamonds denote operations conditioned on the spin in the $y$-basis. The $XBS$ and $U_{BS}$ operations are driven by different pairs of beams that need not be phase-locked to each other, since the phase $\Phi$ can be arbitrary.}
\label{circuit diagram}
\end{figure*}

Our demonstration utilizes three motional modes of two $^{171}\text{Yb}^+$ ions confined in a linear Paul trap. Fig.~\ref{circuit diagram} shows the gate and measurement sequence. The modes $a$ and $b$ are the axial in-phase (IP) and out-of-phase (OP) modes, with secular frequencies \(2\pi\times\) \{0.591, 1.01\} MHz, respectively. The auxiliary mode $c$ is a radial OP mode with secular frequency \(2\pi\times\) 1.16 MHz and is initialized in the motional ground state. The spin states are encoded in ground-state hyperfine levels \(\ket{g}=\ket{^2S_{1/2},F=0,m_F=0}\) and \(\ket{e}=\ket{^2S_{1/2},F=1,m_F=0}\) of one of the \(^{171}\text{Yb}^+\) ions, which are separated by 12.643 GHz and magnetically insensitive to first order. Before the experiment, the other ion is shelved into a metastable dark $^2F_{7/2}$ state by a 398.9786 nm laser that drives the transition $^2D_{3/2} \rightarrow{}^{3}[1/2]_{3/2}$, and does not interact with the laser beams during the experiment. Spin control and spin-motion couplings are implemented using two-photon stimulated Raman transitions \cite{hayes2010entanglement}, driven by mode-locked  pulsed laser beams with a pulse duration of 3 ps, a central wavelength of 372 nm, and a repetition rate of 75.974 MHz. We initialize the
spin and motional states using optical pumping, electromagnetically induced transparency (EIT) and resolved sideband cooling~\cite{RevModPhys.75.281,evers2004double,semerikov2018eit}. The desired motional input states are then prepared using spin-motion couplings~\cite{ding2017quantum,hayes2010entanglement} before applying the eSWAP gate sequence. The spin state is read out via state-dependent fluorescence at the end of the sequence~\cite{olmschenk2007manipulation}. 

We implement the spin-independent beam splitter operation~\cite{fontbote2026error} on trapped ions using the ac Stark shift induced by a pair of \(\sigma^\pm\)-polarized Raman beams at 372 nm detuned from \(^2S_{1/2} \leftrightarrow ^2P_{1/2}\) transition.  For a frequency difference \(\Delta \omega_L\) between the beams, the Stark shift can be approximated by
\begin{align}
    \delta_{ac}(t) = \frac{d^2E^2}{6\hbar^2\Delta}\cos(\Delta\vec{k}\cdot\vec{r}-\Delta\omega_L t)
~,
\label{stark shift}
\end{align}
where \(E\) is the electric field amplitude of each beam, $d$ is the hyperfine reduced dipole matrix element, \(\Delta\) is the laser detuning from the optical transition, and \(\Delta\vec{k}\) is the effective wavevector. Because the laser detuning, \(|\Delta|/2\pi\approx5.5\) THz, is much larger than the hyperfine splitting, the modulation is approximately equal for \(\ket{g}\) and \(\ket{e}\), which results in the interaction Hamiltonian,
$H_{ss} \approx \hbar\Omega_{ss} \mathds{1}\otimes\exp\qty(i(\Delta\vec{k}\cdot\vec{r}  - \Delta \omega_L t+ \Phi)) +h.c.$, 
where \(\Omega_{ss}\) is the coupling strength, proportional to the ac Stark shift modulation amplitude, and $\Phi$ is the phase difference of the two Raman beams. The residual differential Stark shift is compensated by a spin-echo sequence, where the spin is flipped midway through the operation to absorb the residual Stark shift into a global phase offset. Therefore, driving \(\Delta\omega_L=\omega_b-\omega_c\) yields the spin-independent
beam splitter Hamiltonian,
\begin{align}
H_{BS} &= -\hbar\eta_b\eta_c\Omega_{ss} \qty(\hat{b}^{\dagger} \hat{c} e^{i\Phi} + \hat{b} \hat{c} ^{\dagger}e^{-i\Phi})
~,
\label{eq:H_BS}
\end{align}
where \(\eta_b\), \(\omega_b\) and \(\hat{b}^{\dagger}\) (\(\eta_c\), \(\omega_c\) and \(\hat{c}^{\dagger}\)) are the Lamb-Dicke parameter, mode frequency and creation operator of the mode $b$ ($c$), respectively. The corresponding time evolution is the beam splitter operator, $U_{BS}(\Theta,\Phi)  = \exp(\frac{i}{2}\Theta\qty(\hat{b}^{\dagger} \hat{c} e^{i\Phi} + \hat{b} \hat{c} ^{\dagger}e^{-i\Phi}))$, where \(\Theta=2\eta_b\eta_c\Omega_{ss}t\).

At $\Theta = \pm\pi$, the beam splitter fully exchanges the populations of the two modes, but differs from the SWAP operator by a phase factor that depends on the laser phase $\Phi$ and the motional mode states. If the two modes are prepared in Fock states, the transformation is
\begin{align}
U_{BS}(\pm\pi,\Phi) |j,k\rangle 
&=(\pm i)^{j+k}e^{i(k-j)\Phi}SWAP\ket{j,k}\nonumber\\
&=(\pm i)^{j+k}e^{i(k-j)\Phi}\ket{k,j}
~.
\end{align}

Extending these operations to an eSWAP requires control over the motional state-dependent phases that accompany the exchange.

In this work, as depicted in Fig. \ref{circuit diagram}, we achieve this using the spin-independent, optically driven $U_{BS}$ operation to transfer the unwanted phase from state exchange to an empty auxiliary motional mode without introducing additional spin-motion entanglement.

State exchange is realized using spin-dependent beam splitter interactions, allowing an auxiliary-spin rotation to control the SWAP amplitudes. To implement a spin-dependent beam splitter interaction~\cite{PhysRevLett.124.170502,nguyen_experimental_2021,hou2024coherent}, we drive second-order red and blue motional sidebands using bi-chromatic beams that are detuned \(\pm(\omega_a-\omega_b)\) from the carrier frequency with phases $\Phi_1$ and $\Phi_2$~\cite{nguyen_experimental_2021,PhysRevLett.124.170502}. The Hamiltonian for the spin-dependent beam splitter interaction is
\begin{align}
    H_{SBS}&=-\hbar\eta_a\eta_b\Omega ( \hat{\sigma}_+{\hat{a}^{\dagger}} \hat{b}e^{i\Phi_1}
    +\hat{\sigma}_+{\hat{a}} \hat{b}^{\dagger}e^{i\Phi_2}) +h.c.
    \label{eq:H_SBS}
\end{align}
where modes $a$ and $b$ are the modes of interest. With \(\Theta=2\eta_a\eta_b\Omega t\), the spin-dependent beam splitter operator is $U_{BS}(\Theta(\cos\Phi_s\hat{\sigma}_x+\sin\Phi_s\hat{\sigma}_y),\Phi_m)$, where $\Phi_s = (\Phi_1+\Phi_2)/2$ and $\Phi_m = (\Phi_1-\Phi_2)/2$ are the spin phase and motional mode phase, respectively.

Setting $\Phi_s=0$ and $\Phi_m = \frac{\pi}{2}$ defines the $x$-basis conditional beam splitter ($XBS$):
\begin{align}
    XBS(\Theta) \equiv U_{BS}\qty(\Theta\hat{\sigma}_x,\frac{\pi}{2})
    ~.
    \label{XBS}
\end{align}

Similarly, setting $\Phi_s=\frac{\pi}{2}$ gives the $y$-basis conditional beam splitter ($YBS$). Inserting a spin rotation between a pair of $YBS$ gates yields the exponential-parity gate ($eP$), a central component of the eSWAP gate sequence:
\begin{align}
eP(\theta)&\equiv YBS(-\pi)R_x(\theta)YBS(\pi)\nonumber\\
&=\mathds{1}\cos\frac{\theta}{2}-i\hat{\sigma}_x(-1)^{\hat{n}_a+\hat{n}_b}\sin\frac{\theta}{2}
~,
\label{EP}
\end{align}
where \(\hat{n}_a\) and \(\hat{n}_b\) are the number operators of modes $a$ and $b$. 

Combining $U_{BS}(\pm\pi,\Phi)$, $XBS(\pm\frac{\pi}{2})$ and $eP(\theta)$ gates gives the eSWAP gate as shown in Fig. \ref{circuit diagram}, where $\Phi$ can be arbitrary (see Supplemental Material), 
 \begin{align} \label{eswap}
     \text{eSWAP}(\theta) 
     &= \mathds{1}\cos\frac{\theta}{2}-i\text{SWAP}\sin\frac{\theta}{2}
     ~,
 \end{align}
provided that the spin is initialized in \(\ket{+}=\frac{1}{\sqrt{2}}(\ket{g}+\ket{e})\) and mode $c$ is initially in the ground state. 

To characterize the eSWAP operation and probe residual motional state-dependent phases, we measure selected slices of the joint characteristic function~\cite{millican_engineering_2025,qd1c-1fk9,fluhmann_direct_2020} as
\begin{align}
\chi(\alpha,\beta)=\Tr[\rho_{ab}D_a\left(\alpha\right)D_b\left(\beta\right)]
~,
\end{align}
where $D_a(\alpha)$ and $D_b(\beta)$ are displacement operators on modes $a$ and $b$, respectively. In principle, the full four-dimensional joint characteristic function contains complete information about the two-mode state~\cite{fluhmann_direct_2020,millican_engineering_2025,jeon2025two}. However, acquiring the full four-dimensional function would require substantially longer measurement times. Here, our aim is to test if the gate sequence produces the required identity-SWAP superposition in Eq.~(\ref{eswap}) rather than reconstruct the complete two-mode density matrix. We therefore measure selected characteristic-function slices instead of performing full multimode tomography.

We adopt the direct two-mode characteristic-function measurement scheme~\cite{fluhmann_direct_2020,millican_engineering_2025}, as shown on the right side of Fig. \ref{circuit diagram}. Spin-dependent displacements, \(D(\hat{\sigma}_y \alpha )\), are driven by simultaneous first-order red and blue sidebands using bi-chromatic beams detuned from the carrier frequency by the corresponding mode frequencies, \(\pm\omega_i\),
\begin{align}
    H&=i\hbar\eta\Omega\hat{\sigma}_y\qty(\hat{a}^\dagger e^{i\Phi_m}-\hat{a}e^{-i\Phi_m})
    \label{eq:H_D}
    ~.
\end{align}

Starting from the spin state \(|g\rangle\), we apply a spin rotation $R_y(\phi)$, followed by spin-dependent displacements \(D_a\qty(\hat{\sigma}_y \frac{\alpha}{2} )\) and \(D_b\qty(\hat{\sigma}_y \frac{\beta}{2} )\) on modes $a$ and $b$, respectively. The expectation value of \(\hat{\sigma}_z\) yields
\begin{align}
    \expval{ \hat{\sigma}_z}
&=\cos\phi \Re[\chi(\alpha,\beta)]+  
\sin\phi \Im[\chi(\alpha,\beta)]
~,
\end{align}
where the choice of $\phi=0$ or $\frac{\pi}{2}$ selects the real or imaginary part of the joint characteristic function, respectively.

Before the characteristic function measurement sequence, a spin rotation followed by optical pumping resets the spin from its eSWAP
initialization state $\ket{+}$ to $\ket{g}$. We infer the value of $\langle \hat{\sigma}_z\rangle$ from the
measured spin excited-state population using \(\langle \hat{\sigma}_z \rangle=1-2P(|e\rangle\langle e|)\).

\begin{figure}[t!]
\makebox[\linewidth][c]{%
    \hspace{-0.0\linewidth}%
    \includegraphics[
    trim={30mm 0mm 15mm 15mm},
    clip,
    width=1.05\linewidth]{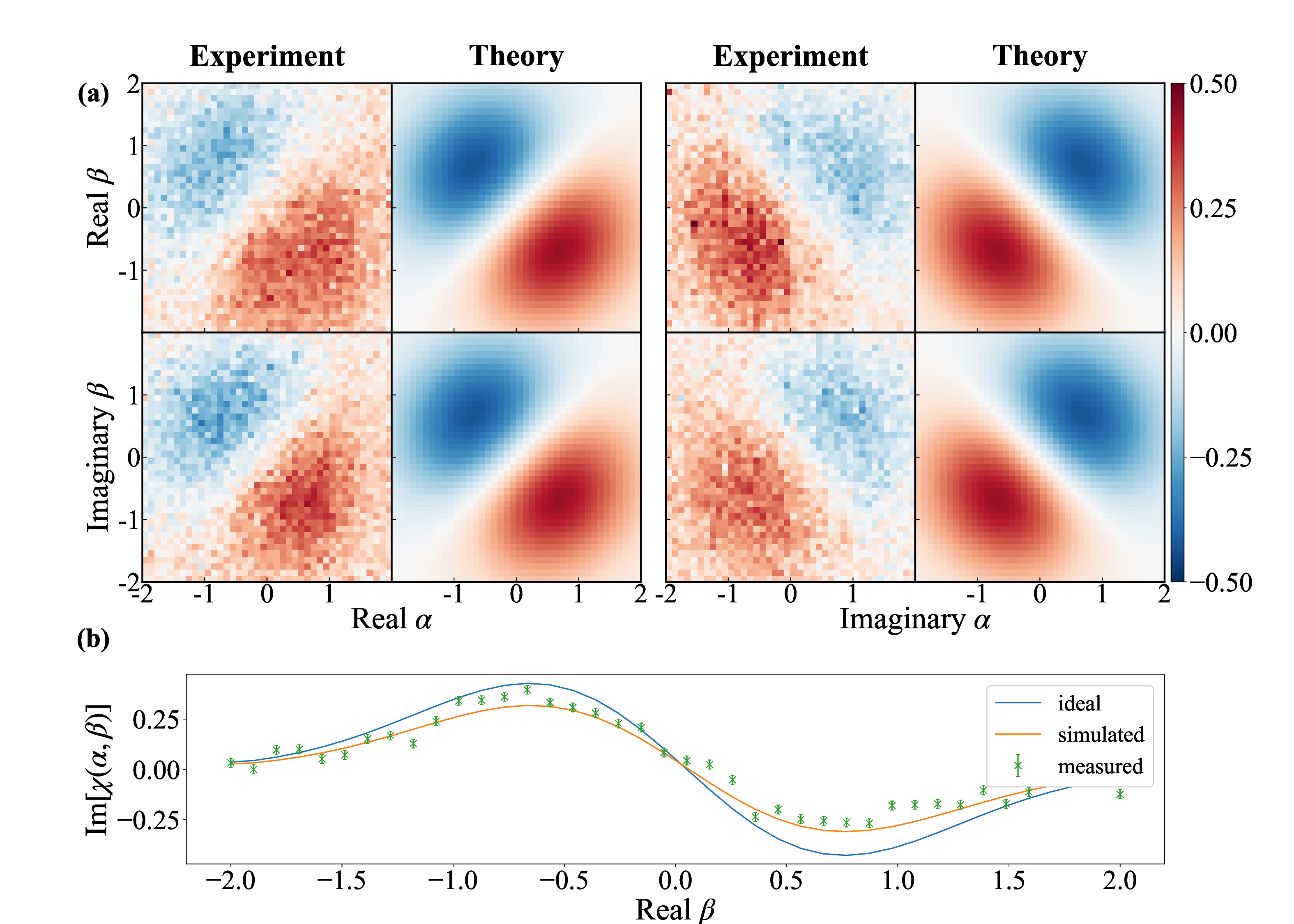}%
}
\setlength{\abovecaptionskip}{-8pt}
\caption{\label{chara_0_0+1} (a) Imaginary slices of the joint characteristic function of \((\ket{0}(\ket{0}+\ket{1}) + i (\ket{0}+\ket{1})\ket{0})/2\). Each experimentally measured 40 $\times$ 40 slice contains data points averaged over 500 experimental repetitions. The data are compared with the prediction for an ideal \(\text{eSWAP}(-\frac{\pi}{2})\) gate. (b) Measured line cut $\Re(\alpha)=-\Re(\beta)$, compared with the ideal prediction and a numerical simulation. The error bars are plotted according to
the binomial statistics of measurements.}
\end{figure}

To benchmark the performance of the $\text{eSWAP}(-\pi/2)$ gate, we first prepare a product state $\ket{\psi_a, \phi_b} = \frac{1}{\sqrt{2}}\ket{0}(\ket{0}+\ket{1})$, which ideally evolves into the entangled state $\ket{\Psi_{\text{out}}} = \frac{1}{2} [ \ket{0}(\ket{0}+\ket{1}) + i(\ket{0}+\ket{1})\ket{0} ]$. This state is sensitive to both the relative phase of the identity-SWAP superposition and the relative phase of Fock-state components. These phases produce distinguishable characteristic rotations of the imaginary characteristic-function slices (see Supplemental Material), as shown in Fig. \ref{chara_0_0+1}(a).

To quantify the agreement between the measured and ideal characteristic function, we calculate the overlap for each slice of the experimental and simulated data as
\begin{align}
\mathcal{O}_i=\frac{\sum_{\alpha,\beta}\chi_{ideal}(\alpha,\beta)\chi_{i}(\alpha,\beta)}{\sum_{\alpha,\beta}\chi_{ideal}^2(\alpha,\beta)}, \quad
i\in\{\mathrm{exp},\mathrm{sim}\} 
~.
\end{align}

The four slices yield an average overlap of $\mathcal{O}_{exp}=(63\pm 4)\%$. The reduced contrast of the measured characteristic function can be attributed to decoherence and errors accumulated during the composite sequence. To investigate this effect, we simulate the complete sequence using the QuTiP Lindblad master-equation solver~\cite{qutip5} independently incorporating the measured heating and dephasing rates of all three motional modes. The simulations identify dephasing of the auxiliary radial OP mode, whose measured coherence time is approximately $(10\pm 2)$ ms, as the dominant contribution. In comparison, both axial modes have coherence times of approximately 100 ms or longer, while heating of the IP mode at $(17\pm 1)$ phonons/s produces only a negligible effect over the 4.2 ms sequence. The simulation reproduces the contrast reduction observed in Fig. \ref{chara_0_0+1}(b), with $\mathcal{O}_{sim} = (75\pm3)\%$. The remaining discrepancy may partly arise from imperfections not included in the simulations, such as state-detection errors estimated to reduce the overlap by approximately $2\%$ and imperfect initialization of the motional states. 

\begin{figure}[h]
\makebox[\linewidth][c]{%
    \hspace{-0.0\linewidth}%
    \includegraphics[
    trim={35mm 25mm 15mm 15mm},
    clip,
    width=1.1\linewidth]{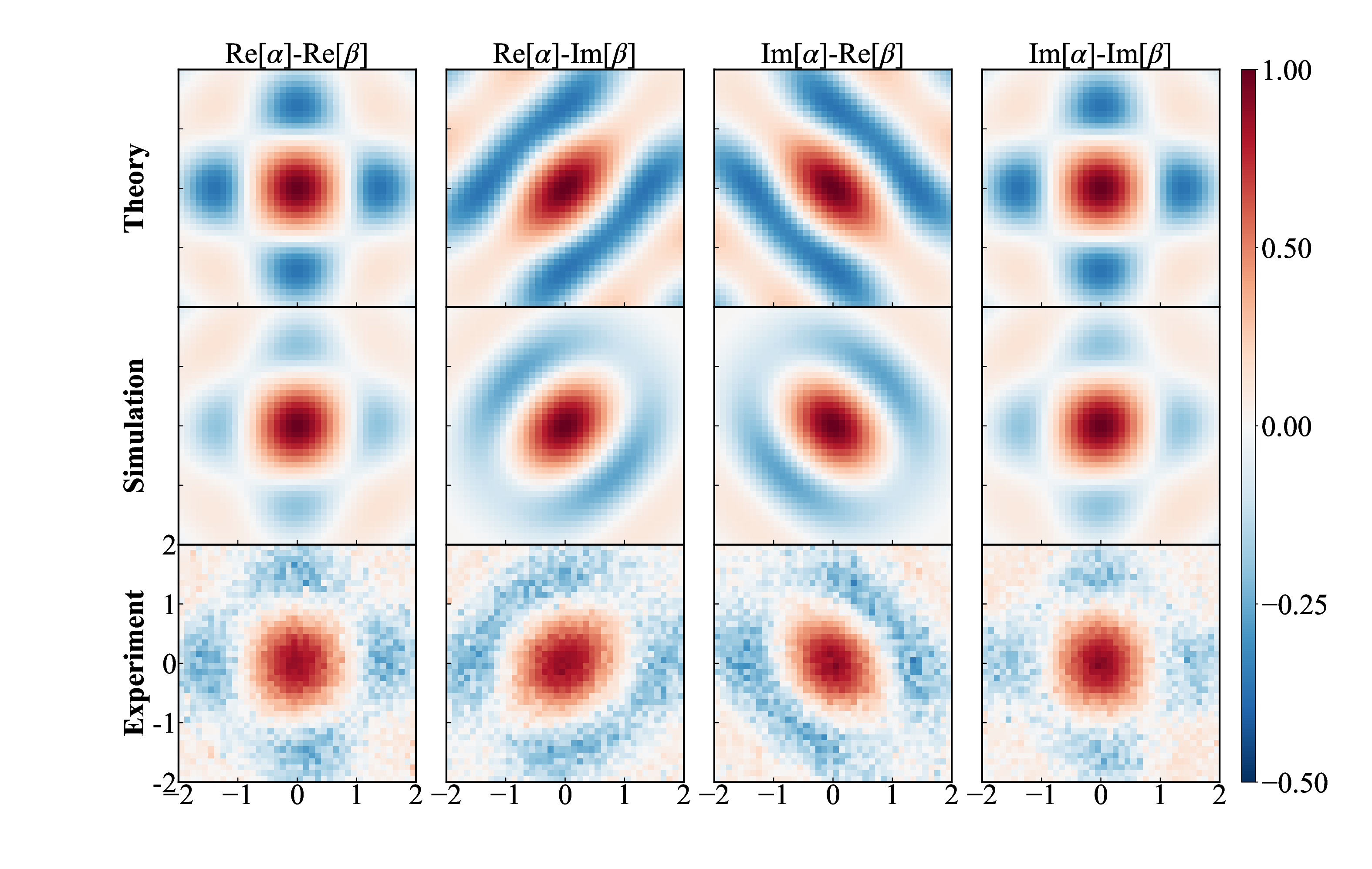}%
}
\caption{\label{chara_1_2}Real slices of the joint characteristic function of \((\ket{1}\ket{2} + i \ket{2}\ket{1})/\sqrt{2}\). Each experimentally measured 40 $\times$ 40 slice contains data points averaged over 500 experimental repetitions. The data are compared with the ideal \(\text{eSWAP}(-\frac{\pi}{2})\) prediction and a numerical simulation of auxiliary-mode dephasing.}
\end{figure}

The ideal gate sequence shown in Fig. \ref{circuit diagram} compensates the motional state-dependent phases associated with beam splitter state transfer for arbitrary bosonic inputs. To test this compensation beyond the lowest Fock states, we apply $\text{eSWAP}(-\pi/2)$ to $\ket{\psi_a, \varphi_b} = \ket{1, 2}$, for which both motional modes are excited and have opposite parity. The target state, $\ket{\Psi_{\text{out}}} = \frac{1}{\sqrt{2}} (\ket{1}\ket{2} + i\ket{2}\ket{1})$, has a purely real joint characteristic function with characteristic symmetries in the selected two-dimensional slices (see Supplemental Material). Therefore, we measured four representative real slices, shown in Fig. \ref{chara_1_2}. The four slices yield an average overlap of $\mathcal{O}_{exp}=(78\pm3)\%$, compared to \(\mathcal{O}_{sim} = (81\pm2)\%\). The simulation also reproduces the observed distortion in the characteristic-function patterns. The characteristic function features, including their orientations in the Re[$\alpha$]-Im[$\beta$] and Im[$\alpha$]-Re[$\beta$] planes, reproduce the expected relative phase between the $\ket{1, 2}$ and $\ket{2, 1}$ components. Together with the measurements in Fig. \ref{chara_0_0+1}, these results
demonstrate phase compensation across both odd and even motional occupations, consistent with the input-state-independent eSWAP operation of Eq. (\ref{eswap}). 

\begin{figure}
\vspace{0.04\linewidth}%
\makebox[\linewidth][c]{%
    \hspace{-0.02\linewidth}%
    \includegraphics[
    trim={10mm 5mm 10mm 0mm},
    clip,
    width=1.05\linewidth]{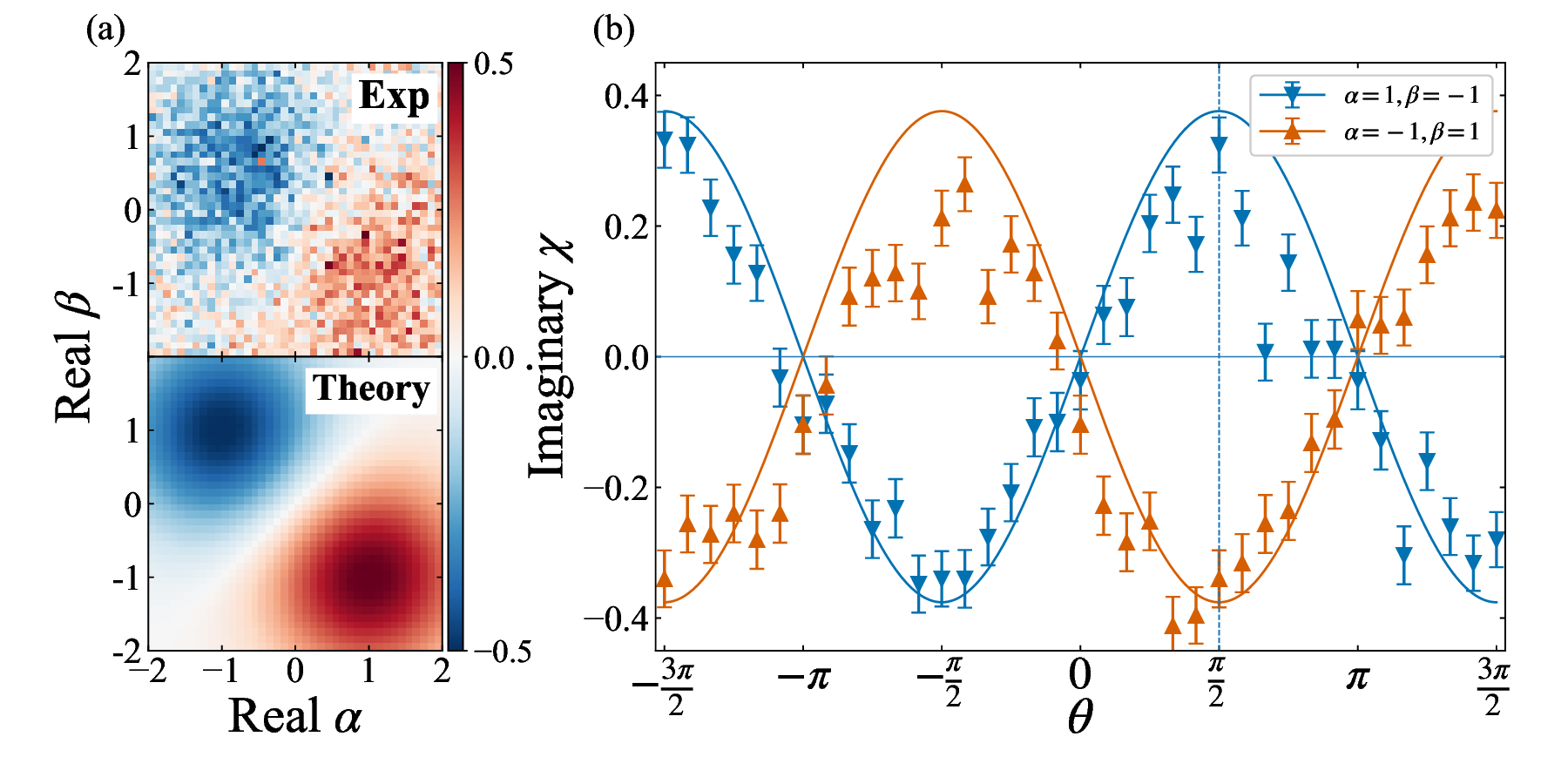}%
}
\setlength{\abovecaptionskip}{-8pt}
\caption{\label{oscillation}(a) Ideal and measured $\Re({\alpha})$-$\Re(\beta)$ slices of \(\Im[\chi(\alpha,\beta)]\) at $\theta=\frac{\pi}{2}$ over 100 experimental repetitions. (b) Measured (triangles) and simulated (curves) values of $\Im{[\chi(\pm2\gamma,\mp2\gamma)]}$ as a function of $\theta$, showing the expected $2\pi$ periodicity. Each data point is measured over 500 experimental repetitions and the error bars are plotted according to the binomial statistics of measurements.}
\end{figure}

Finally, to verify continuous control of the amplitudes in Eq.~(\ref{eswap}), we prepared the coherent state input \(\ket{\psi_a, \phi_b} = \ket{-\gamma, \gamma}\) for $\gamma=0.5$. Applying $\text{eSWAP}(\theta)$ ideally produces the two-mode entangled cat state $\ket{\Psi_{\text{out}}} = \cos\frac{\theta}{2} \ket{-\gamma, \gamma} - i\sin\frac{\theta}{2}\ket{\gamma, -\gamma}$, whose characteristic function has an imaginary part satisfying \(\Im{[\chi(\pm2\gamma,\mp2\gamma)]}\approx\pm0.5\sin\theta\). In contrast to the two states considered above, numerical simulations indicate that dephasing of the auxiliary radial OP mode has a negligible effect on this signal, while heating of the axial IP mode reduces the oscillation amplitude to approximately 0.37. As shown in Fig. \ref{oscillation}, the measured characteristic-function values follow the expected sinusoidal dependence on $\theta$ and agree with the simulation, confirming continuous coherent control of the relative amplitudes in Eq.~(\ref{eswap}).

In conclusion, we have demonstrated the first experimental realization of an eSWAP gate between bosonic motional modes of trapped ions, by combining spin-independent and spin-dependent beam splitter operations. Selected slices of the joint characteristic function are consistent with the predicted input-state-independent eSWAP($-\frac{\pi}{2}$) operation for the tested distinct input states. Measurements as a function of the eSWAP angle demonstrate coherent control of the amplitudes in Eq.~(\ref{eswap}). This construction can also be adapted to other hybrid platforms. The coherent, tunable exchange provided by the eSWAP gate makes it a useful primitive for hybrid quantum information processing, bosonic quantum error correction, and quantum simulation. The present performance is limited by auxiliary-mode dephasing and motional heating; reducing these errors would improve the gate performance. 

\section*{Acknowledgements}
This work is supported by the National Research Foundation, Singapore through the National Quantum Office, hosted in A*STAR, under its Centre for Quantum Technologies Funding Initiative (S24Q2d0009) and Quantum Engineering Programme (W21Qpd0208). The authors would like to thank J. You and N.B.J. Lee for their constructive feedback and assistance and Yvonne Gao for her helpful suggestions.
The authors used ChatGPT (OpenAI; GPT-5.6 and GPT-6) to assist with proofreading of the manuscript.
\bibliography{references}

@article{wineland1998experimental,
  title={Experimental issues in coherent quantum-state manipulation of trapped atomic ions},
  author={Wineland, David J and Monroe, Christopher and Itano, Wayne M and Leibfried, Dietrich and King, Brian E and Meekhof, Dawn M},
  journal={Journal of research of the National Institute of Standards and Technology},
  volume={103},
  number={3},
  pages={259},
  year={1998}
}

@article{haffner2008quantum,
  title = {Quantum computing with trapped ions},
  author = {H{\"a}ffner, H. and Roos, C. F. and Blatt, R.},
  journal = {Physics Reports},
  volume = {469},
  number = {4},
  pages = {155--203},
  year = {2008},
  publisher = {Elsevier},
  doi = {10.1016/j.physrep.2008.09.003}
}

@article{lau_universal_2016,
	title = {Universal {Quantum} {Computing} with {Arbitrary} {Continuous}-{Variable} {Encoding}},
	volume = {117},
	copyright = {http://link.aps.org/licenses/aps-default-license},
	issn = {0031-9007, 1079-7114},
	url = {https://link.aps.org/doi/10.1103/PhysRevLett.117.100501},
	doi = {10.1103/PhysRevLett.117.100501},
	number = {10},
	urldate = {2026-03-03},
	journal = {Physical Review Letters},
	author = {Lau, Hoi-Kwan and Plenio, Martin B.},
	month = aug,
	year = {2016},
	pages = {100501},
}

@article{gao_entanglement_2019,
	title = {Entanglement of bosonic modes through an engineered exchange interaction},
	volume = {566},
	copyright = {2019 The Author(s), under exclusive licence to Springer Nature Limited},
	issn = {1476-4687},
	url = {https://www.nature.com/articles/s41586-019-0970-4},
	doi = {10.1038/s41586-019-0970-4},
	number = {7745},
	urldate = {2026-02-28},
	journal = {Nature},
	publisher = {Nature Publishing Group},
	author = {Gao, Yvonne Y. and Lester, Brian J. and Chou, Kevin S. and Frunzio, Luigi and Devoret, Michel H. and Jiang, Liang and Girvin, S. M. and Schoelkopf, Robert J.},
	month = feb,
	year = {2019},
	pages = {509--512},
}

@article{nguyen2021quantum,
  title={Quantum-enhanced bosonic learning machine},
  author={Nguyen, Chi-Huan and Tseng, Ko-Wei and Maslennikov, Gleb and Gan, HCJ and Matsukevich, Dzmitry},
  journal={arXiv preprint arXiv:2104.04168},
  year={2021}
}

@article{nguyen_experimental_2021,
  title={Experimental SWAP test of infinite dimensional quantum states},
  author={Nguyen, Chi-Huan and Tseng, Ko-Wei and Maslennikov, Gleb and Gan, HCJ and Matsukevich, Dzmitry},
  journal={arXiv preprint arXiv:2103.10219},
  year={2021}
}

@article{fluhmann_direct_2020,
  title={Direct characteristic-function tomography of quantum states of the trapped-ion motional oscillator},
  author={Fl{\"u}hmann, Christa and Home, Jonathan P},
  journal={Physical Review Letters},
  volume={125},
  number={4},
  pages={043602},
  year={2020},
  publisher={APS}
}

@article{millican_engineering_2025,
  title={Engineering continuous-variable entanglement in mechanical oscillators with optimal control},
  author={Millican, Maverick J and Matsos, Vassili G and Valahu, Christophe H and Navickas, Tomas and Bond, Liam J and Tan, Ting Rei},
  journal={Physical Review Letters},
  volume={135},
  number={23},
  pages={233604},
  year={2025},
  publisher={APS}
}

@article{RevModPhys.77.513,
  title = {Quantum information with continuous variables},
  author = {Braunstein, Samuel L. and van Loock, Peter}, 
  journal = {Rev. Mod. Phys.},
  volume = {77},
  issue = {2},
  pages = {513--577},
  numpages = {0},
  year = {2005},
  month = {Jun},
  publisher = {American Physical Society},
  doi = {10.1103/RevModPhys.77.513},
  url = {https://link.aps.org/doi/10.1103/RevModPhys.77.513}
}

@article{kemper2025hybrid,
  title={Hybrid continuous-discrete-variable quantum computing: a guide to utility},
  author={Kemper, AF and Alvertis, Antonios and Asaduzzaman, Muhammad and Bakalov, Bojko N and Baron, Dror and Bierman, Joel and Burgstahler, Blake and Chundury, Srikar and Das, Elin Ranjan and Furches, Jim and others},
  journal={arXiv preprint arXiv:2511.13882},
  year={2025}
}

@article{HybridPRXQuantum,
  title = {Hybrid Oscillator-Qubit Quantum Processors: Instruction Set Architectures, Abstract Machine Models, and Applications},
  author = {Liu, Yuan and Singh, Shraddha and Smith, Kevin C. and Crane, Eleanor and Martyn, John M. and Eickbusch, Alec and Schuckert, Alexander and Li, Richard D. and Sinanan-Singh, Jasmine and Soley, Micheline B. and Tsunoda, Takahiro and Chuang, Isaac L. and Wiebe, Nathan and Girvin, Steven M.},
  journal = {PRX Quantum},
  volume = {7},
  issue = {1},
  pages = {010201},
  numpages = {166},
  year = {2026},
  month = {Jan},
  publisher = {American Physical Society},
  doi = {10.1103/4rf7-9tfx},
  url = {https://link.aps.org/doi/10.1103/4rf7-9tfx}
}

@article{qd1c-1fk9,
  title = {Optimal Fidelity Estimation from Binary Measurements for Discrete and Continuous Variable Systems},
  author = {Fawzi, Omar and Oufkir, Aadil and Salzmann, Robert},
  journal = {PRX Quantum},
  volume = {7},
  issue = {1},
  pages = {010309},
  numpages = {32},
  year = {2026},
  month = {Jan},
  publisher = {American Physical Society},
  doi = {10.1103/qd1c-1fk9},
  url = {https://link.aps.org/doi/10.1103/qd1c-1fk9}
}

@article{chen_quantum_2021,
	title = {Quantum computation and simulation with vibrational modes of trapped ions},
	volume = {30},
	issn = {1674-1056},
	url = {http://arxiv.org/abs/2103.14299},
	doi = {10.1088/1674-1056/ac01e3},
	number = {6},
	urldate = {2026-07-20},
	journal = {Chinese Physics B},
	author = {Chen, Wentao and Gan, Jaren and Zhang, Jing-Ning and Matsukevich, Dzmitry and Kim, Kihwan},
	month = jun,
	year = {2021},
	note = {arXiv:2103.14299 [quant-ph]},
	pages = {060311},
}

@article{lloyd_quantum_2014,
	title = {Quantum principal component analysis},
	volume = {10},
	copyright = {2014 Springer Nature Limited},
	issn = {1745-2481},
	url = {https://www.nature.com/articles/nphys3029},
	doi = {10.1038/nphys3029},
	number = {9},
	urldate = {2026-07-20},
	journal = {Nature Physics},
	publisher = {Nature Publishing Group},
	author = {Lloyd, Seth and Mohseni, Masoud and Rebentrost, Patrick},
	month = sep,
	year = {2014},
	pages = {631--633},
}

@article{jeon2025two,
  title={Two-mode bosonic state tomography with single-shot joint-parity measurement of a trapped ion},
  author={Jeon, Honggi and Kang, Jiyong and Choi, Wonhyeong and Kim, Kyunghye and You, Jaehun and Kim, Taehyun},
  journal={PRX Quantum},
  volume={6},
  number={4},
  pages={040352},
  year={2025},
  publisher={APS}
}

@article{lau_quantum_2017,
	title = {Quantum {Machine} {Learning} over {Infinite} {Dimensions}},
	volume = {118},
	issn = {0031-9007, 1079-7114},
	url = {https://link.aps.org/doi/10.1103/PhysRevLett.118.080501},
	doi = {10.1103/PhysRevLett.118.080501},
	number = {8},
	urldate = {2022-08-29},
	journal = {Physical Review Letters},
	author = {Lau, Hoi-Kwan and Pooser, Raphael and Siopsis, George and Weedbrook, Christian},
	month = feb,
	year = {2017},
	pages = {080501},
}

@article{PhysRevLett.124.170502,
  title = {Hybrid Quantum Computing with Conditional Beam Splitter Gate in Trapped Ion System},
  author = {Gan, H. C. J. and Maslennikov, Gleb and Tseng, Ko-Wei and Nguyen, Chihuan and Matsukevich, Dzmitry},
  journal = {Phys. Rev. Lett.},
  volume = {124},
  issue = {17},
  pages = {170502},
  numpages = {6},
  year = {2020},
  month = {Apr},
  publisher = {American Physical Society},
  doi = {10.1103/PhysRevLett.124.170502},
  url = {https://link.aps.org/doi/10.1103/PhysRevLett.124.170502}
}

@article{hayes2010entanglement,
  title = {Entanglement of Atomic Qubits Using an Optical Frequency Comb},
  author = {Hayes, D. and Matsukevich, D. N. and Maunz, P. and Hucul, D. and Quraishi, Q. and Olmschenk, S. and Campbell, W. and Mizrahi, J. and Senko, C. and Monroe, C.},
  journal = {Phys. Rev. Lett.},
  volume = {104},
  issue = {14},
  pages = {140501},
  numpages = {4},
  year = {2010},
  month = {Apr},
  publisher = {American Physical Society},
  doi = {10.1103/PhysRevLett.104.140501},
  url = {https://link.aps.org/doi/10.1103/PhysRevLett.104.140501}
}

@article{fluhmann2019encoding,
  title={Encoding a qubit in a trapped-ion mechanical oscillator},
  author={Fl{\"u}hmann, Christa and Nguyen, Thanh Long and Marinelli, Matteo and Negnevitsky, Vlad and Mehta, Karan and Home, Jonathan P},
  journal={Nature},
  volume={566},
  number={7745},
  pages={513--517},
  year={2019},
  publisher={Nature Publishing Group UK London}
}

@article{kienzler2015quantum,
  title={Quantum harmonic oscillator state synthesis by reservoir engineering},
  author={Kienzler, Daniel and Lo, H-Y and Keitch, B and De Clercq, L and Leupold, F and Lindenfelser, F and Marinelli, M and Negnevitsky, Vlad and Home, JP},
  journal={Science},
  volume={347},
  number={6217},
  pages={53--56},
  year={2015},
  publisher={American Association for the Advancement of Science}
}

@article{chen2023scalable,
  title={Scalable and programmable phononic network with trapped ions},
  author={Chen, Wentao and Lu, Yao and Zhang, Shuaining and Zhang, Kuan and Huang, Guanhao and Qiao, Mu and Su, Xiaolu and Zhang, Jialiang and Zhang, Jing-Ning and Banchi, Leonardo and others},
  journal={Nature Physics},
  volume={19},
  number={6},
  pages={877--883},
  year={2023},
  publisher={Nature Publishing Group UK London}
}

@article{tu_precision_2025,
  title={Precision polarization tuning for light shift mitigation in trapped-ion qubits},
  author={Tu, Hengchao and Luan, Chun-Yang and Zou, Menglin and Yin, Zihan and Rehan, Kamran and Kim, Kihwan},
  journal={Physical Review Research},
  volume={7},
  number={4},
  pages={043002},
  year={2025},
  publisher={APS}
}

@article{RevModPhys.75.281,
  title = {Quantum dynamics of single trapped ions},
  author = {Leibfried, D. and Blatt, R. and Monroe, C. and Wineland, D.},
  journal = {Rev. Mod. Phys.},
  volume = {75},
  issue = {1},
  pages = {281--324},
  numpages = {0},
  year = {2003},
  month = {Mar},
  publisher = {American Physical Society},
  doi = {10.1103/RevModPhys.75.281},
  url = {https://link.aps.org/doi/10.1103/RevModPhys.75.281}
}

@article{fontbote2026error,
  title={Error Correction of Beamsplitter-Generated Entangled {GKP} States},
  author={Fontbot{\'e}-Schmidt, Moritz and Metzner, Jeremy and Berterotti{\`e}re, Florence and Rojkov, Ivan and Ferk, Alexander and Stadler, Martin and D{\"o}nmez, Bahadir and Berner, Ralf and Welte, Stephan and Kienzler, Daniel and others},
  journal={arXiv preprint arXiv:2605.08009},
  year={2026}
}

@article{PhysRevLett.82.1784,
  title = {Quantum Computation over Continuous Variables},
  author = {Lloyd, Seth and Braunstein, Samuel L.},
  journal = {Phys. Rev. Lett.},
  volume = {82},
  issue = {8},
  pages = {1784--1787},
  numpages = {0},
  year = {1999},
  month = {Feb},
  publisher = {American Physical Society},
  doi = {10.1103/PhysRevLett.82.1784},
  url = {https://link.aps.org/doi/10.1103/PhysRevLett.82.1784}
}

@article{van2011optical,
  title={Optical hybrid approaches to quantum information},
  author={van Loock, Peter},
  journal={Laser \& Photonics Reviews},
  volume={5},
  number={2},
  pages={167--200},
  year={2011},
  publisher={Wiley Online Library}
}

@phdthesis{egan_scaling_2021,
  title={Scaling quantum computers with long chains of trapped ions},
  author={Egan, Laird Nicholas},
  year={2021},
  school={University of Maryland, College Park}
}

@article{qutip5,
  title={{QuTiP} 5: The quantum toolbox in {Python}},
  author={Lambert, Neill and Gigu{\`e}re, Eric and Menczel, Paul and Li, Boxi and Hopf, Patrick and Su{\'a}rez, Gerardo and Gali, Marc and Lishman, Jake and Gadhvi, Rushiraj and Agarwal, Rochisha and others},
  journal={Physics Reports},
  volume={1153},
  pages={1--62},
  year={2026},
  publisher={Elsevier}
}

@article{cahill1969ordered,
  title={Ordered expansions in boson amplitude operators},
  author={Cahill, Kevin E and Glauber, Roy J},
  journal={Physical Review},
  volume={177},
  number={5},
  pages={1857},
  year={1969},
  publisher={APS}
}

@article{hou2024coherent,
  title={Coherent coupling and non-destructive measurement of trapped-ion mechanical oscillators},
  author={Hou, Pan-Yu and Wu, Jenny J and Erickson, Stephen D and Cole, Daniel C and Zarantonello, Giorgio and Brandt, Adam D and Geller, Shawn and Kwiatkowski, Alex and Glancy, Scott and Knill, Emanuel and others},
  journal={Nature Physics},
  volume={20},
  number={10},
  pages={1636--1641},
  year={2024},
  publisher={Nature Publishing Group UK London}
}

@article{sutherland2021universal,
  title={Universal hybrid quantum computing in trapped ions},
  author={Sutherland, RT and Srinivas, R},
  journal={Physical Review A},
  volume={104},
  number={3},
  pages={032609},
  year={2021},
  publisher={APS}
}

@article{michael2016new,
  title={New class of quantum error-correcting codes for a bosonic mode},
  author={Michael, Marios H and Silveri, Matti and Brierley, RT and Albert, Victor V and Salmilehto, Juha and Jiang, Liang and Girvin, Steven M},
  journal={Physical Review X},
  volume={6},
  number={3},
  pages={031006},
  year={2016},
  publisher={APS}
}

@article{ofek2016extending,
  title={Extending the lifetime of a quantum bit with error correction in superconducting circuits},
  author={Ofek, Nissim and Petrenko, Andrei and Heeres, Reinier and Reinhold, Philip and Leghtas, Zaki and Vlastakis, Brian and Liu, Yehan and Frunzio, Luigi and Girvin, Steven M and Jiang, Liang and others},
  journal={Nature},
  volume={536},
  number={7617},
  pages={441--445},
  year={2016},
  publisher={Nature Publishing Group UK London}
}

@article{gottesman2000encoding,
  title = {Encoding a qubit in an oscillator},
  author = {Gottesman, Daniel and Kitaev, Alexei and Preskill, John},
  journal = {Phys. Rev. A},
  volume = {64},
  issue = {1},
  pages = {012310},
  numpages = {21},
  year = {2001},
  month = {Jun},
  publisher = {American Physical Society},
  doi = {10.1103/PhysRevA.64.012310},
  url = {https://link.aps.org/doi/10.1103/PhysRevA.64.012310}
}

@article{putterman2025hardware,
  title={Hardware-efficient quantum error correction via concatenated bosonic qubits},
  author={Putterman, Harald and Noh, Kyungjoo and Hann, Connor T and MacCabe, Gregory S and Aghaeimeibodi, Shahriar and Patel, Rishi N and Lee, Menyoung and Jones, William M and Moradinejad, Hesam and Rodriguez, Roberto and others},
  journal={Nature},
  volume={638},
  number={8052},
  pages={927--934},
  year={2025},
  publisher={Nature Publishing Group UK London}
}

@article{katz2023programmable,
  title={Programmable quantum simulations of bosonic systems with trapped ions},
  author={Katz, Or and Monroe, Christopher},
  journal={Physical Review Letters},
  volume={131},
  number={3},
  pages={033604},
  year={2023},
  publisher={APS}
}

@article{olmschenk2007manipulation,
  title={Manipulation and detection of a trapped $\text{Yb}^+$ hyperfine qubit},
  author={Olmschenk, Steve and Younge, Kelly C and Moehring, David L and Matsukevich, Dzmitry N and Maunz, Peter and Monroe, Christopher},
  journal={Physical Review A—Atomic, Molecular, and Optical Physics},
  volume={76},
  number={5},
  pages={052314},
  year={2007},
  publisher={APS}
}

@article{evers2004double,
  title={Double-{EIT} ground-state laser cooling without blue-sideband heating},
  author={Evers, J{\"o}rg and Keitel, Christoph H},
  journal={EPL (Europhysics Letters)},
  volume={68},
  number={3},
  pages={370--376},
  year={2004}
}

@article{semerikov2018eit,
  title={{EIT} ground state cooling scheme of $^{171}\text{Yb}^+$ based on the $^2{S}_{1/2}\rightarrow ^2{P}_{1/2}$ cooling transition},
  author={Semerikov, IA and Zalivako, IV and Borisenko, AS and Khabarova, KY and Kolachevsky, NN},
  journal={Journal of Russian Laser Research},
  volume={39},
  number={6},
  pages={568--574},
  year={2018},
  publisher={Springer}
}

@article{ding2017quantum,
  title={Quantum parametric oscillator with trapped ions},
  author={Ding, Shiqian and Maslennikov, Gleb and Habl{\"u}tzel, Roland and Loh, Huanqian and Matsukevich, Dzmitry},
  journal={Physical Review Letters},
  volume={119},
  number={15},
  pages={150404},
  year={2017},
  publisher={APS}
}

@article{campagne2020quantum,
  title={Quantum error correction of a qubit encoded in grid states of an oscillator},
  author={Campagne-Ibarcq, Philippe and Eickbusch, Alec and Touzard, Steven and Zalys-Geller, Evan and Frattini, Nicholas E and Sivak, Volodymyr V and Reinhold, Philip and Puri, Shruti and Shankar, Shyam and Schoelkopf, Robert J and others},
  journal={Nature},
  volume={584},
  number={7821},
  pages={368--372},
  year={2020},
  publisher={Nature Publishing Group UK London}
}

@article{de2022error,
  title={Error correction of a logical grid state qubit by dissipative pumping},
  author={de Neeve, Brennan and Nguyen, Thanh-Long and Behrle, Tanja and Home, Jonathan P},
  journal={Nature Physics},
  volume={18},
  number={3},
  pages={296--300},
  year={2022},
  publisher={Nature Publishing Group UK London}
}

@article{slim2024optomechanical,
  title={Optomechanical realization of the bosonic {Kitaev} chain},
  author={Slim, Jesse J and Wanjura, Clara C and Brunelli, Matteo and del Pino, Javier and Nunnenkamp, Andreas and Verhagen, Ewold},
  journal={Nature},
  volume={627},
  number={8005},
  pages={767--771},
  year={2024},
  publisher={Nature Publishing Group UK London}
}

@article{varona2024towards,
  title={Towards quantum computing {Feynman} diagrams in hybrid qubit-oscillator devices},
  author={Varona, S and Saner, S and B{\u{a}}z{\u{a}}van, O and Araneda, G and Aarts, G and Bermudez, A},
  journal={arXiv preprint arXiv:2411.05092},
  year={2024}
}

@article{zhang2018experimental,
  title={Experimental quantum simulation of fermion-antifermion scattering via boson exchange in a trapped ion},
  author={Zhang, Xiang and Zhang, Kuan and Shen, Yangchao and Zhang, Shuaining and Zhang, Jing-Ning and Yung, Man-Hong and Casanova, Jorge and Pedernales, Julen S and Lamata, Lucas and Solano, Enrique and others},
  journal={Nature Communications},
  volume={9},
  number={1},
  pages={195},
  year={2018},
  publisher={Nature Publishing Group UK London}
}

@article{jastrzebski2026adaptive,
  title={Adaptive Relational Learning on Multi-instance Quantum Data with Photonic Processors},
  author={Jastrzebski, Marcin and Yu, Shang and Patel, Raj B and Kyriienko, Oleksandr},
  journal={arXiv preprint arXiv:2609.17352},
  year={2026}
}

@article{zhong2020quantum,
  title={Quantum computational advantage using photons},
  author={Zhong, Han-Sen and Wang, Hui and Deng, Yu-Hao and Chen, Ming-Cheng and Peng, Li-Chao and Luo, Yi-Han and Qin, Jian and Wu, Dian and Ding, Xing and Hu, Yi and others},
  journal={Science},
  volume={370},
  number={6523},
  pages={1460--1463},
  year={2020},
  publisher={American Association for the Advancement of Science}
}

@article{madsen2022quantum,
  title={Quantum computational advantage with a programmable photonic processor},
  author={Madsen, Lars S and Laudenbach, Fabian and Askarani, Mohsen Falamarzi and Rortais, Fabien and Vincent, Trevor and Bulmer, Jacob FF and Miatto, Filippo M and Neuhaus, Leonhard and Helt, Lukas G and Collins, Matthew J and others},
  journal={Nature},
  volume={606},
  number={7912},
  pages={75--81},
  year={2022},
  publisher={Nature Publishing Group UK London}
}

@article{asavanant2019generation,
  title={Generation of time-domain-multiplexed two-dimensional cluster state},
  author={Asavanant, Warit and Shiozawa, Yu and Yokoyama, Shota and Charoensombutamon, Baramee and Emura, Hiroki and Alexander, Rafael N and Takeda, Shuntaro and Yoshikawa, Jun-ichi and Menicucci, Nicolas C and Yonezawa, Hidehiro and others},
  journal={Science},
  volume={366},
  number={6463},
  pages={373--376},
  year={2019},
  publisher={American Association for the Advancement of Science}
}

@article{konno2024logical,
  title={Logical states for fault-tolerant quantum computation with propagating light},
  author={Konno, Shunya and Asavanant, Warit and Hanamura, Fumiya and Nagayoshi, Hironari and Fukui, Kosuke and Sakaguchi, Atsushi and Ide, Ryuhoh and China, Fumihiro and Yabuno, Masahiro and Miki, Shigehito and others},
  journal={Science},
  volume={383},
  number={6680},
  pages={289--293},
  year={2024},
  publisher={American Association for the Advancement of Science}
}

@article{larsen2025integrated,
  title={Integrated photonic source of {Gottesman--Kitaev--Preskill} qubits},
  author={Larsen, Mikkel V and Bourassa, J Eli and Kocsis, Sacha and Tasker, Joel F and Chadwick, Robert S and Gonz{\'a}lez-Arciniegas, Carlos and Hastrup, Jacob and Lopetegui-Gonz{\'a}lez, Carlos E and Miatto, Filippo M and Motamedi, A and others},
  journal={Nature},
  volume={642},
  number={8068},
  pages={587--591},
  year={2025},
  publisher={Nature Publishing Group UK London}
}

@article{shaw2025erasure,
  title={Erasure cooling, control, and hyperentanglement of motion in optical tweezers},
  author={Shaw, Adam L and Scholl, Pascal and Finkelstein, Ran and Tsai, Richard Bing-Shiun and Choi, Joonhee and Endres, Manuel},
  journal={Science},
  volume={388},
  number={6749},
  pages={845--849},
  year={2025},
  publisher={American Association for the Advancement of Science}
}

@article{brown2023time,
  title={Time-of-flight quantum tomography of an atom in an optical tweezer},
  author={Brown, MO and Muleady, SR and Dworschack, WJ and Lewis-Swan, RJ and Rey, AM and Romero-Isart, O and Regal, CA},
  journal={Nature Physics},
  volume={19},
  number={4},
  pages={569--573},
  year={2023},
  publisher={Nature Publishing Group UK London}
}

@article{hartke2022quantum,
  title={Quantum register of fermion pairs},
  author={Hartke, Thomas and Oreg, Botond and Jia, Ningyuan and Zwierlein, Martin},
  journal={Nature},
  volume={601},
  number={7894},
  pages={537--541},
  year={2022},
  publisher={Nature Publishing Group UK London}
}

@article{lienhard2025generation,
  title={Generation of motional squeezed states for neutral atoms in optical tweezers},
  author={Lienhard, Vincent and Martin, Romain and Chew, Yuki Torii and Tomita, Takafumi and Ohmori, Kenji and de L{\'e}s{\'e}leuc, Sylvain},
  journal={Physical Review Letters},
  volume={135},
  number={25},
  pages={253404},
  year={2025},
  publisher={APS}
}

@article{eickbusch2022fast,
  title={Fast universal control of an oscillator with weak dispersive coupling to a qubit},
  author={Eickbusch, Alec and Sivak, Volodymyr and Ding, Andy Z and Elder, Salvatore S and Jha, Shantanu R and Venkatraman, Jayameenakshi and Royer, Baptiste and Girvin, Steven M and Schoelkopf, Robert J and Devoret, Michel H},
  journal={Nature Physics},
  volume={18},
  number={12},
  pages={1464--1469},
  year={2022},
  publisher={Nature Publishing Group UK London}
}

@article{hu2019quantum,
  title={Quantum error correction and universal gate set operation on a binomial bosonic logical qubit},
  author={Hu, Ling and Ma, Yuwei and Cai, Weizhou and Mu, Xianghao and Xu, Yuan and Wang, Weiting and Wu, Yukai and Wang, Haiyan and Song, YP and Zou, C-L and others},
  journal={Nature Physics},
  volume={15},
  number={5},
  pages={503--508},
  year={2019},
  publisher={Nature Publishing Group UK London}
}

@article{sivak2023real,
  title={Real-time quantum error correction beyond break-even},
  author={Sivak, Volodymyr V and Eickbusch, Alec and Royer, Baptiste and Singh, Shraddha and Tsioutsios, Ioannis and Ganjam, Suhas and Miano, Alessandro and Brock, Benjamin L and Ding, Andy Z and Frunzio, Luigi and others},
  journal={Nature},
  volume={616},
  number={7955},
  pages={50--55},
  year={2023},
  publisher={Nature Publishing Group UK London}
}

@article{brock2025quantum,
  title={Quantum error correction of qudits beyond break-even},
  author={Brock, Benjamin L and Singh, Shraddha and Eickbusch, Alec and Sivak, Volodymyr V and Ding, Andy Z and Frunzio, Luigi and Girvin, Steven M and Devoret, Michel H},
  journal={Nature},
  volume={641},
  number={8063},
  pages={612--618},
  year={2025},
  publisher={Nature Publishing Group UK London}
}

@article{sutherland2022one,
  title={One- and two-qubit gate infidelities due to motional errors in trapped ions and electrons},
  author={Sutherland, R Tyler and Yu, Qian and Beck, Kristin M and H{\"a}ffner, Hartmut},
  journal={Physical Review A},
  volume={105},
  number={2},
  pages={022437},
  year={2022},
  publisher={APS}
}

@article{saner2026generating,
  title={Generating arbitrary superpositions of nonclassical quantum harmonic oscillator states},
  author={Saner, S and B{\u{a}}z{\u{a}}van, O and Webb, DJ and Araneda, G and Lucas, DM and Ballance, CJ and Srinivas, R},
  journal={Physical Review X},
  volume={16},
  number={2},
  pages={021049},
  year={2026},
  publisher={APS}
}

@article{PhysRevLett.123.080602,
  title = {Spin Heat Engine Coupled to a Harmonic-Oscillator Flywheel},
  author = {von Lindenfels, D. and Gr{\"a}b, O. and Schmiegelow, C. T. and Kaushal, V. and Schulz, J. and Mitchison, Mark T. and Goold, John and Schmidt-Kaler, F. and Poschinger, U. G.},
  journal = {Phys. Rev. Lett.},
  volume = {123},
  issue = {8},
  pages = {080602},
  numpages = {6},
  year = {2019},
  month = {Aug},
  publisher = {American Physical Society},
  doi = {10.1103/PhysRevLett.123.080602},
  url = {https://link.aps.org/doi/10.1103/PhysRevLett.123.080602}
}

@article{maslennikov2019quantum,
  title={Quantum absorption refrigerator with trapped ions},
  author={Maslennikov, Gleb and Ding, Shiqian and Habl{\"u}tzel, Roland and Gan, Jaren and Roulet, Alexandre and Nimmrichter, Stefan and Dai, Jibo and Scarani, Valerio and Matsukevich, Dzmitry},
  journal={Nature Communications},
  volume={10},
  number={1},
  pages={202},
  year={2019},
  publisher={Nature Publishing Group UK London}
}

@article{heeres2015cavity,
  title={Cavity state manipulation using photon-number selective phase gates},
  author={Heeres, Reinier W and Vlastakis, Brian and Holland, Eric and Krastanov, Stefan and Albert, Victor V and Frunzio, Luigi and Jiang, Liang and Schoelkopf, Robert J},
  journal={Physical Review Letters},
  volume={115},
  number={13},
  pages={137002},
  year={2015},
  publisher={APS}
}

@article{wang2016schrodinger,
  title={A {Schr{\"o}dinger} cat living in two boxes},
  author={Wang, Chen and Gao, Yvonne Y and Reinhold, Philip and Heeres, Reinier W and Ofek, Nissim and Chou, Kevin and Axline, Christopher and Reagor, Matthew and Blumoff, Jacob and Sliwa, KM and others},
  journal={Science},
  volume={352},
  number={6289},
  pages={1087--1091},
  year={2016},
  publisher={American Association for the Advancement of Science}
}

@article{gao2018programmable,
  title={Programmable interference between two microwave quantum memories},
  author={Gao, Yvonne Y and Lester, Brian J and Zhang, Yaxing and Wang, Chen and Rosenblum, Serge and Frunzio, Luigi and Jiang, Liang and Girvin, SM and Schoelkopf, Robert J},
  journal={Physical Review X},
  volume={8},
  number={2},
  pages={021073},
  year={2018},
  publisher={APS}
}

@article{jarlaud2021coherence,
  title={Coherence properties of highly-excited motional states of a trapped ion},
  author={Jarlaud, Vincent and Hrmo, Pavel and Joshi, Manoj K and Thompson, Richard C},
  journal={Journal of Physics B: Atomic, Molecular and Optical Physics},
  volume={54},
  number={1},
  pages={015501},
  year={2021},
  publisher={IOP Publishing}
}

@article{lucas2007long,
  title={A long-lived memory qubit on a low-decoherence quantum bus},
  author={Lucas, DM and Keitch, BC and Home, JP and Imreh, G and McDonnell, MJ and Stacey, DN and Szwer, DJ and Steane, AM},
  journal={arXiv preprint arXiv:0710.4421},
  year={2007}
}

@misc{liu2025robustquantumcomputationaladvantage,
      title={Robust quantum computational advantage with programmable 3050-photon {Gaussian} boson sampling}, 
      author={Hua-Liang Liu and Hao Su and Si-Qiu Gong and Yi-Chao Gu and Hao-Yang Tang and Meng-Hao Jia and Qian Wei and Yukun Song and Dongzhou Wang and Mingyang Zheng and Faxi Chen and Libo Li and Siyu Ren and Xuezhi Zhu and Meihong Wang and Yaojian Chen and Yanfei Liu and Longsheng Song and Pengyu Yang and Junshi Chen and Hong An and Lei Zhang and Lin Gan and Guangwen Yang and Jia-Min Xu and Yu-Ming He and Hui Wang and Han-Sen Zhong and Ming-Cheng Chen and Xiao Jiang and Li Li and Nai-Le Liu and Yu-Hao Deng and Xiao-Long Su and Qiang Zhang and Chao-Yang Lu and Jian-Wei Pan},
      year={2025},
      eprint={2508.09092},
      archivePrefix={arXiv},
      primaryClass={quant-ph},
      url={https://arxiv.org/abs/2508.09092}, 
}

\clearpage 
\onecolumngrid 
\makeatletter
\counterwithout{equation}{section} 
\numberwithin{equation}{subsection}
\renewcommand{\theequation}{\thesubsection.\arabic{equation}}
\numberwithin{figure}{subsection}
\setlength\parindent{0pt}
\widetext
\begin{center}
\textbf{\large Supplemental Material for: Exponential-SWAP Gate between Bosonic Motional Modes of Trapped Ions}
\end{center}

\subsection{Stark Shift}
Consider a pair of counter-propagating laser beams directed at an ion along its quantization axis. Their total electric field is
\begin{align}
    \vec{E}(t)
    &=\frac{1}{2}\sum_{i=1}^2 \vec{\epsilon}_iE_i\exp\qty(-i(\vec{k}_i\cdot\vec{r}-\omega_it))+c.c.
\end{align}
where $\vec{\epsilon}_i$ is the polarization vector, $E_i$ is the electric-field amplitude, and $\omega_i$ is the frequency. The dipole interaction is
\begin{align}
    H_{dip}(t) = -\hat{d}\cdot\vec{E}(t)=-\frac{1}{2}\sum_{i=1}^2\hat{d}\cdot \vec{\epsilon}_i\mathcal{E}_i +h.c.  ~,
\end{align}
where $\hat{d}$ is the electric-dipole operator and $\mathcal{E}_i=E_ie^{-i(\vec{k}_i\cdot\vec{r}-\omega_it)}$ is the complex field amplitude. The coupling of $\ket{g} = \ket{^2S_{1/2}, F=0, m_F = 0}$ to the $^2P_{1/2}$ manifold of an $^{171}\text{Yb}^+$ ion produces the Stark shift as
\begin{align}
    \delta_{stark}(t) = \frac{1}{4\hbar^2\Delta}\sum_{F,m_F}{\qty|\sum_{i=1}^2\mathcal{E}_i\bra{F, m_F}\hat{d}\cdot \vec{\epsilon}_i\ket{g}|^2} ~,
\end{align}
where $\Delta$ is the laser detuning from the transition between $\ket{g}$ and the $^2P_{1/2}$ manifold. For beams propagating along the quantization axis, only the $\hat{\sigma}_\pm-$transitions contribute. When both beams have the same $\hat{\sigma}_+$ or $\hat{\sigma}_-$ polarization relative to the quantization axis, interference of two beams yields a modulated Stark shift given under the rotating-wave approximation by
\begin{align}
        \delta_{stark}(t) \approx  \frac{d^2}{12\hbar^2\Delta}\qty[|E_1|^2+|E_2|^2+2E_1E_2\cos(\Delta\vec{k}\cdot\vec{r}-\Delta\omega_L t)]~,
\end{align}
where $\Delta \vec{k} = \vec{k}_1-\vec{k}_2$, $\Delta \omega_L = \omega_1-\omega_2$. The polarization-resolved dipole matrix element is $|\bra{^2P_{1/2},F=1,m_F=\pm1}\hat{d} \cdot\hat{\sigma}^\pm \ket{g}|^2=\frac{1}{3}d^2$, where $d$ is the hyperfine reduced dipole matrix element. The first two terms are static while the final term corresponds to the ac Stark shift modulation in space and time. For equal field amplitudes,  $E_1=E_2=E$, the last term is
\begin{align}
    \delta_{ac}(t) = \frac{d^2E^2}{6\hbar^2\Delta}\cos(\Delta\vec{k}\cdot\vec{r}-\Delta\omega_L t)~.
\end{align}
The ac Stark shift of $\ket{e}=\ket{^2S_{1/2}, F=1, m_F = 0}$ due to coupling to the $^2P_{1/2}$ manifold can be derived in the same way. 

\subsection{Beam Splitter Interactions}
The beam splitter interaction used in this work is
\begin{align}
U_{BS}(\theta,\phi) = \exp\left(i\frac{\theta}{2}(\hat{a}^\dagger \hat{b}e^{i\phi}+\hat{a}\hat{b}^\dagger e^{-i\phi})\right) ~,\label{eq:UBS}
\end{align}
where \(\hat{a}^{\dagger}\) and  \(\hat{b}^{\dagger}\) (\(\hat{a}\) and \(\hat{b}\)) are the creation (annihilation) operators of bosonic modes $a$ and $b$. Expanding the exponential shows that this unitary transforms  \(\hat{a}^{\dagger}\) and \(\hat{b}^{\dagger}\) as follows,
\begin{align}
U_{BS}(\theta,\phi)\hat{a}^{\dagger}U_{BS}^{\dagger}(\theta,\phi)&=\hat{a}^{\dagger}\cos{\frac{\theta}{2}}+\hat{b}^{\dagger}i e^{-i\phi}\sin{\frac{\theta}{2}} ~, \nonumber \\
U_{BS}(\theta,\phi)\hat{b}^{\dagger}U_{BS}^{\dagger}(\theta,\phi)&=\hat{b}^{\dagger}\cos{\frac{\theta}{2}}+\hat{a}^{\dagger}i e^{i\phi}\sin{\frac{\theta}{2}}  ~.
\label{UBS transform}
\end{align}

Using these transformations and the unitarity of \(U_{BS}(\theta,\phi)\), 
the beam splitter operator transforms an arbitrary two-mode Fock state \(|j,k \rangle\) as
\begin{align}
U_{BS}(\theta,\phi)|j,k\rangle = \frac{1}{\sqrt{j!k!}}\left(\hat{a}^\dagger\cos\frac{\theta}{2}+\hat{b}^\dagger i e^{-i\phi}\sin\frac{\theta}{2}\right)^j
\times\left(\hat{b}^\dagger\cos\frac{\theta}{2}+\hat{a}^\dagger i e^{i\phi}\sin\frac{\theta}{2}\right)^k|0,0\rangle
~. \label{A.3}
\end{align}
At $\theta = \pi$, the populations are fully exchanged. However, each Fock state acquires an
additional phase factor,
\begin{align}
U_{BS}(\pi,\phi)|j,k\rangle &= (i)^{j+k} e^{i(k-j)\phi}|k,j\rangle 
~.
\end{align}

\subsection{eSWAP}\label{proof eSWAP}

The eSWAP gate sequence in Fig. \ref{circuit diagram} of the main text is symmetric about the exponential-parity gate ($eP$), where the sequence before the $eP$ is the Hermitian conjugate of the sequence after it. We therefore begin with
Eq. (\ref{EP}) of the main text,
\begin{align}
eP(\theta) &\equiv YBS(-\pi)R_x(\theta)YBS(\pi)\nonumber\\
&=\mathds{1}\cos\frac{\theta}{2}-i\hat{\sigma}_xY{BS}(2\pi)\sin\frac{\theta}{2}\nonumber\\
&=\mathds{1}\cos\frac{\theta}{2}-i\hat{\sigma}_x(-1)^{\hat{n}_a+\hat{n}_b}\sin\frac{\theta}{2}
~,
\end{align}
where \(\hat{n}_a\) and \(\hat{n}_b\) are the number operators for modes $a$ and $b$. Due to the symmetry of the gate sequence, the surrounding operations cancel in the identity contribution of the exponential-parity gate. Therefore, only the second term needs to be considered.
The $U_{BS}$ operations act only on mode $b$ and auxiliary mode $c$, 
\begin{align}
U_{BS}\qty(\pi, \phi)(-1)^{\hat{n}_b}U_{BS}\left(-\pi, \phi\right)=(-1)^{\hat{n}_c}
~.
\end{align}

The $XBS$ operations act on the spin and modes $a$ and $b$. They transform the remaining factor $\hat{\sigma}_x(-1)^{\hat{n}_a}$ according to 
\begin{align}
    XBS\qty(-\frac{\pi}{2})\hat{\sigma}_x(-1)^{\hat{n}_a}XBS\qty(\frac{\pi}{2})&= \hat{\sigma}_xXBS\qty(-\frac{\pi}{2})\exp(i\pi \hat{a}^\dagger \hat{a})XBS\qty(\frac{\pi}{2})\nonumber\\
    &=\hat{\sigma}_x\exp(i\pi XBS\qty(-\frac{\pi}{2})\hat{a}^\dagger XBS\qty(\frac{\pi}{2})XBS\qty(-\frac{\pi}{2})\hat{a}XBS\qty(\frac{\pi}{2}))\nonumber\\
    &=\hat{\sigma}_x\exp(i\frac{\pi}{2}(\hat{a}^\dagger-\hat{\sigma}_x \hat{b}^\dagger)(\hat{a}-\hat{\sigma}_x \hat{b}))\nonumber\\
    &= \hat{\sigma}_x \exp(i\frac{\pi}{2}(\hat{a}^\dagger \hat{a} +\hat{b}^\dagger \hat{b} - \hat{\sigma}_x \hat{a}^\dagger \hat{b} - \hat{\sigma}_x \hat{a}\hat{b}^\dagger))\nonumber\\
    &=\hat{\sigma}_x \exp(i\frac{\pi}{2}(\hat{a}^\dagger \hat{a} +\hat{b}^\dagger \hat{b}))\exp(-i\frac{\pi}{2}\hat{\sigma}_x(\hat{a}^\dagger \hat{b} + \hat{a}\hat{b}^\dagger))\nonumber\\
    &=\hat{\sigma}_xi^{(\hat{a}^\dagger \hat{a} +\hat{b}^\dagger \hat{b})}U_{BS}\qty(-\pi\hat{\sigma}_x, 0)\nonumber\\
    &= \hat{\sigma}_x i^{(\hat{a}^\dagger \hat{a} +\hat{b}^\dagger \hat{b})}(-\hat{\sigma}_x i)^{(\hat{a}^\dagger \hat{a} +\hat{b}^\dagger \hat{b})}\text{SWAP}
    \label{XNX}
    ~.
\end{align}

The transformation to the third line follows from Eq. (\ref{UBS transform}). The factorization in the fifth line is valid because $\hat{a}^\dagger \hat{a}+\hat{b}^\dagger \hat{b}$ commutes with $\hat{a}^\dagger \hat{b}+\hat{a}\hat{b}^\dagger$:
\begin{align}
    \qty[\hat{a}^\dagger \hat{a}+\hat{b}^\dagger \hat{b},\hat{a}^\dagger \hat{b}+\hat{a}\hat{b}^\dagger]&= \qty[\hat{a}^\dagger \hat{a}, \hat{a}^\dagger \hat{b}+\hat{a}\hat{b}^\dagger] + \qty[\hat{b}^\dagger \hat{b},\hat{a}^\dagger \hat{b}+\hat{a}\hat{b}^\dagger]\nonumber\\
    &=\hat{a}^\dagger \hat{b} - \hat{a}\hat{b}^\dagger + \hat{a}\hat{b}^\dagger -\hat{a}^\dagger \hat{b}\nonumber\\
    &= 0
    ~.
\end{align}

For the initial spin state \(\ket{+}=\frac{1}{\sqrt{2}}(\ket{g}+\ket{e})\), Eq. (\ref{XNX}) reduces to
\begin{align}
    XBS\qty(-\frac{\pi}{2})\hat{\sigma}_x(-1)^{\hat{n}_a}XBS\qty(\frac{\pi}{2})\nonumber &= i^{(\hat{a}^\dagger \hat{a} +\hat{b}^\dagger \hat{b})}(- i)^{(\hat{a}^\dagger \hat{a} +\hat{b}^\dagger \hat{b})}\text{SWAP}\\
    &= \text{SWAP}
    ~.
\end{align}

Combining these results gives
\begin{align}
    \text{eSWAP}(\theta)
    &=\mathds{1}\cos\frac{\theta}{2}-i(-1)^{\hat{n}_c}\text{SWAP}\sin\frac{\theta}{2}
    ~.
\end{align}

Initializing the auxiliary mode $c$ in its ground state gives $\hat{n}_c \ket{0}= 0$, yielding the desired gate,
\begin{align}
    \text{eSWAP}(\theta)
    &=\mathds{1}\cos\frac{\theta}{2}-i\text{SWAP}\sin\frac{\theta}{2}
    ~.
\end{align}

This completes the derivation.

\subsection{Calibration and Stark Shift Compensation}
\begin{figure}[h!]
\setkeys{Gin}{width=0.33\linewidth}
\captionsetup[subfloat]{position=top, justification=raggedright, singlelinecheck=false, skip=0pt}
    \subfloat[\label{detuning} ]{\includegraphics[trim=0cm 0cm 0cm 0.3cm, clip,height=4.7cm]{ 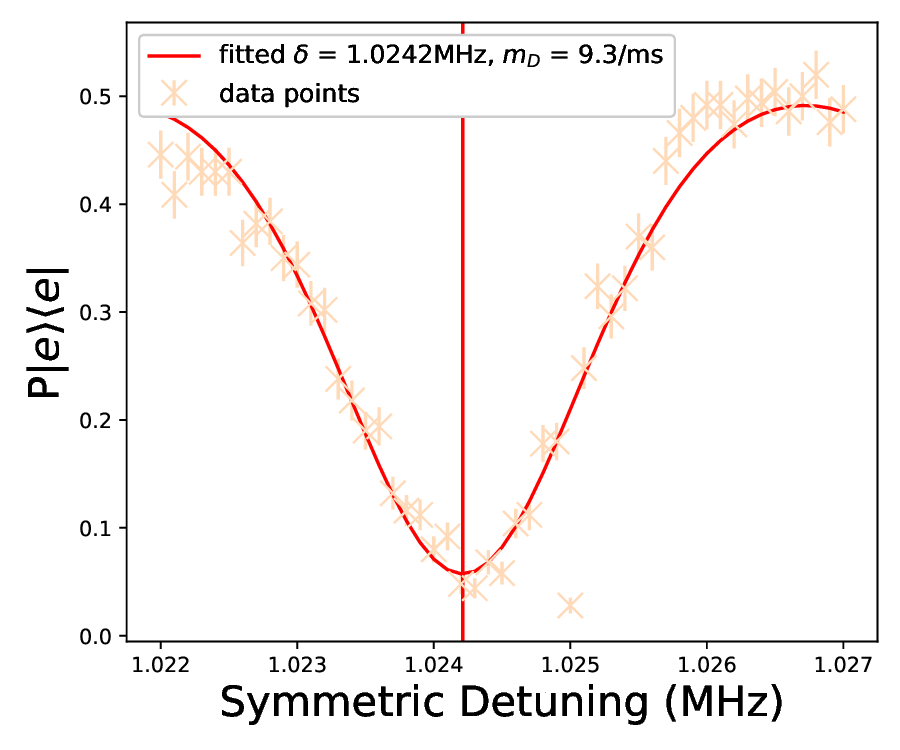}}\hfill%
    \subfloat[\label{amplitude}]{\includegraphics[trim=0cm 0cm 0cm 0.3cm, clip, height=4.7cm]{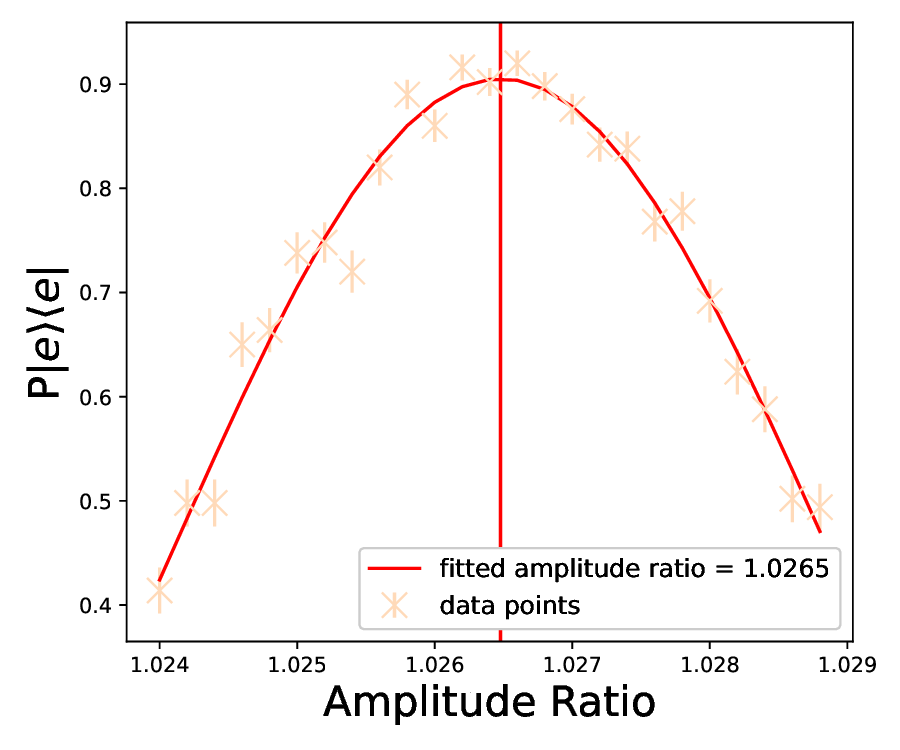}}\hfill%
    \subfloat[\label{gradient}]{{\includegraphics[trim=0cm 0cm 0cm 0.3cm, clip, height=4.7cm]{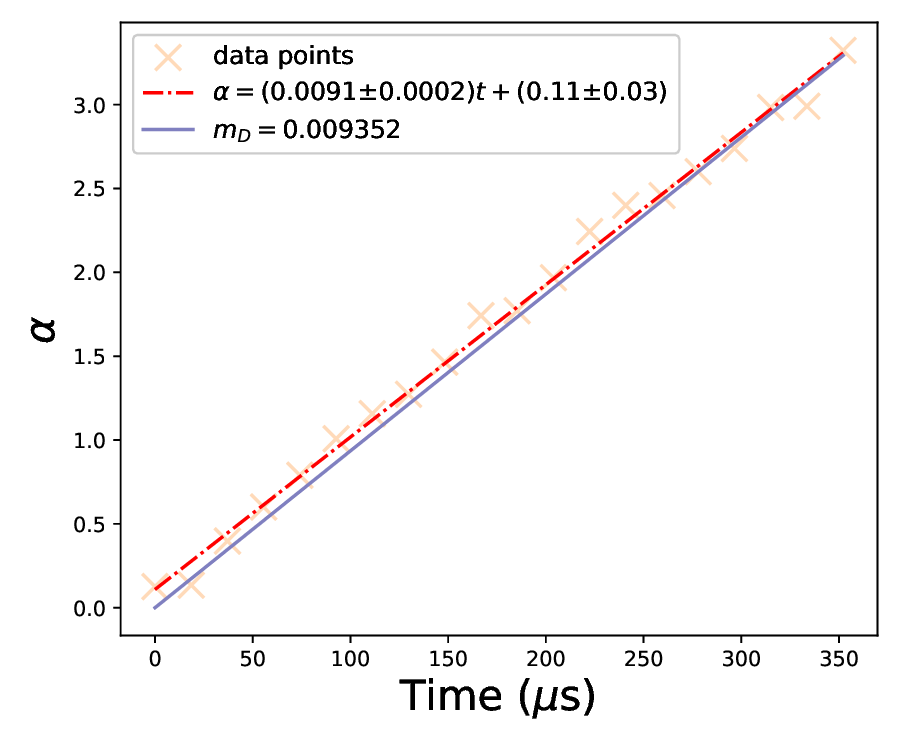}}}
\caption{\label{calibration} (a) Symmetric detuning scan fitted with Eq. (\ref{detuning fit}) including a background offset. (b)  Sideband amplitude ratio scan fitted with a
Gaussian function. Panels (a) and (b) show calibrations for a displacement pulse of duration $t=100$ $\mu$s. (c)  Displacement amplitudes extracted from fits to blue-sideband evolution recorded over 500 $\mu$s. Each data point is averaged over 500 experimental repetitions. The dashed line is a linear fit with a nonzero intercept at 0.11, attributed to imperfect ground-state cooling of the mode and detector dark counts. The solid line shows  $\alpha = m_{D}t$,
using the displacement gradient $m_{D}$ calibrated in (a).}
\end{figure}

The displacement, $XBS$ and $U_{BS}$ operations are calibrated daily prior to the experiment and rechecked every 30 minutes during the experiment to track and compensate slow drifts in mode frequencies on the order of 20 Hz/h for the radial modes. Detuning frequencies are calibrated through a Ramsey-like sequence: a displacement, $XBS$, or $U_{BS}$ pulse of duration $t$ is followed, after a delay, by an identical pulse of opposite phase (reverse pulse). At the correct detuning, the two pulses cancel. The calibration precision improves with longer delay time. In this work, a calibration precision of approximately 1 Hz has been achieved. An example of a displacement detuning scan with no delay between the pulses is shown in Fig. \ref{calibration}(a). The displacement gradient can be calibrated by fitting the equation
\begin{align}
P(\ket{e}\bra{e})=\frac{1}{2}-\frac{1}{2}\exp\qty({-8\qty(\frac{m_{D}}{\delta})^2(\cos\delta t-1)^2})+P_{bg}
~,
\label{detuning fit}
\end{align}
where $t$ is the drive time of displacement and reverse pulses, $\delta$ is the symmetric detuning error, $m_{D}$ is the displacement gradient and $P_{bg}$ is the background offset. The displacement amplitude, $\alpha$, can be calculated by the relation $\alpha = m_Dt$. The displacement gradient calibrated from the symmetric detuning scan is in very good agreement with the data points collected for varying $\alpha$, as shown in Fig. \ref{calibration}(c). A best-fit line through the data points is also plotted for comparison.

For displacement and $XBS$ operations, the bi-chromatic frequencies are symmetric about the carrier frequency. Therefore, the second-order ac Stark shifts cancel out. The fourth-order ac Stark shifts, however, induce an intensity-dependent asymmetric detuning, which we compensate using the methods described in \cite{tu_precision_2025,egan_scaling_2021}. Frequency-dependent efficiency of the acousto-optic modulator (AOM) and small differences in beam pointing can produce a power imbalance between the two frequency components. Since the displacement and $XBS$ pulses used are spin-dependent operations proportional to $\hat{\sigma}_x$ or $\hat{\sigma}_y$, a Ramsey-like sequence can be used to calibrate and correct this power imbalance, where the spin state is prepared in an eigenstate of $\hat{\sigma}_x$ or $\hat{\sigma}_y$ and mapped to spin $\ket{e}$ after the $XBS$ or displacement pulse. The spin remains undisturbed at the optimal amplitude ratio. An example is shown in Fig. \ref{calibration}(b).  The amplitude-ratio calibration precision can also be improved by increasing the pulse duration of the $XBS$ or displacement.

\subsection{\label{graphs} Joint Characteristic Functions}
According to \cite{cahill1969ordered},
\begin{align}\langle m|D(\alpha)|n\rangle = \left(\frac{n!}{m!}\right)^{1/2}\alpha^{m-n}e^{-|\alpha|^2/2}L_n^{(m-n)}(|\alpha|^2),\hspace{5mm} m\geq n
\label{exp D}
\end{align}

where $L_n^{(m-n)}(x)$ is an associated Laguerre polynomial and $m,n\in \mathbb{Z}_{\geq0}$. We will assume that $m\geq n$. The case of $m < n$ can be inferred by replacing $\alpha$ with $-\alpha$, interchanging $m$ and $n$, and taking the complex conjugate:
$\langle m|D(\alpha)|n\rangle = \big[\langle n|D(-\alpha)|m\rangle\big]^*$. Using Eq. (\ref{exp D}), the characteristic function of an entangled state \(\cos\varphi\ket{m,n}+e^{i\theta}\sin\varphi\ket{n,m}\), where $m \neq n$, is

\begin{align}\langle D_a(\alpha)D_b(\beta)\rangle &=\cos^2\varphi\langle m|D(\alpha)|m\rangle_a\langle n|D(\beta)|n\rangle_b+\sin^2\varphi\langle n|D(\alpha)|n\rangle_a\langle m|D(\beta)|m\rangle_b\nonumber\\
&\hspace{5mm}+e^{i\theta}\cos\varphi\sin\varphi\langle m|D(\alpha)|n\rangle_a\langle n|D(\beta)|m\rangle_b+e^{-i\theta}\cos\varphi\sin\varphi\langle n|D(\alpha)|m\rangle_a\langle m|D(\beta)|n\rangle_b\nonumber\\
&=e^{-(|\alpha|^2+|\beta|^2)/2} \bigg[\cos^2\varphi L_m^0(|\alpha|^2)L_n^0(|\beta|^2)+\sin^2\varphi L_n^0(|\alpha|^2)L_m^0(|\beta|^2)\nonumber\\
&\hspace{5mm}+\cos\varphi\sin\varphi\left(\frac{n!}{m!}\right)L_n^{(m-n)}(|\alpha|^2)L_n^{(m-n)}(|\beta|^2)\left(e^{i\theta}(-\alpha\beta^*)^{m-n}+e^{-i\theta}(-\alpha^*\beta)^{m-n}\right)\bigg]
\label{chara_equation}
\end{align}

\begin{figure*}[!h]
    \centering
    \includegraphics[
    trim={45mm 20mm 0mm 35mm},
    clip,
    width=0.9\linewidth]{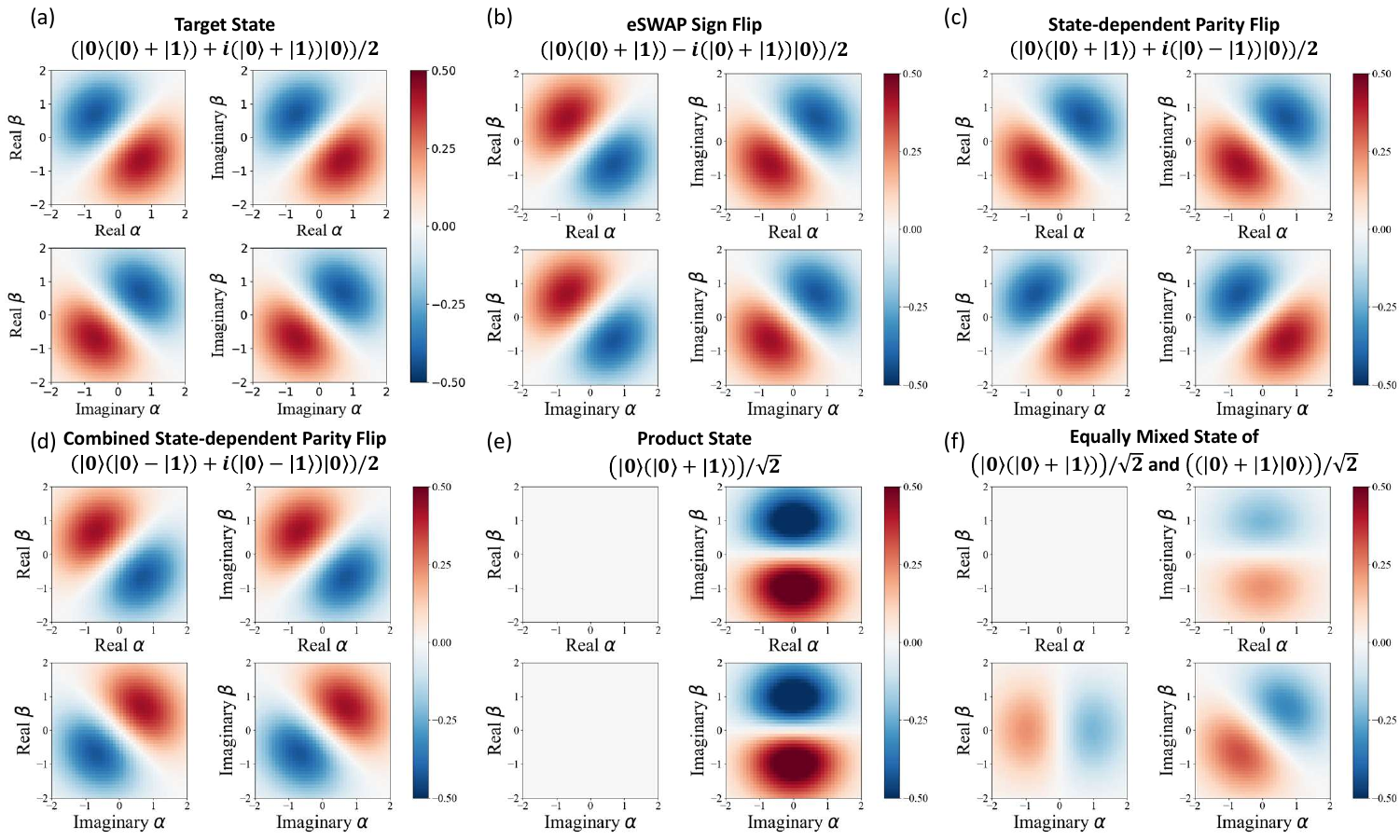}
    \caption{\label{panel_0_0+1}Comparison of the imaginary part of the joint characteristic function for the desired target state \((|0\rangle(|0\rangle+|1\rangle)+i(|0\rangle+|1\rangle)|0\rangle)/2\), representative phase errors, a product state and an incoherent mixture.}
    \includegraphics[
    trim={130mm 20mm 0mm 35mm},
    clip,
    width=0.7\linewidth]{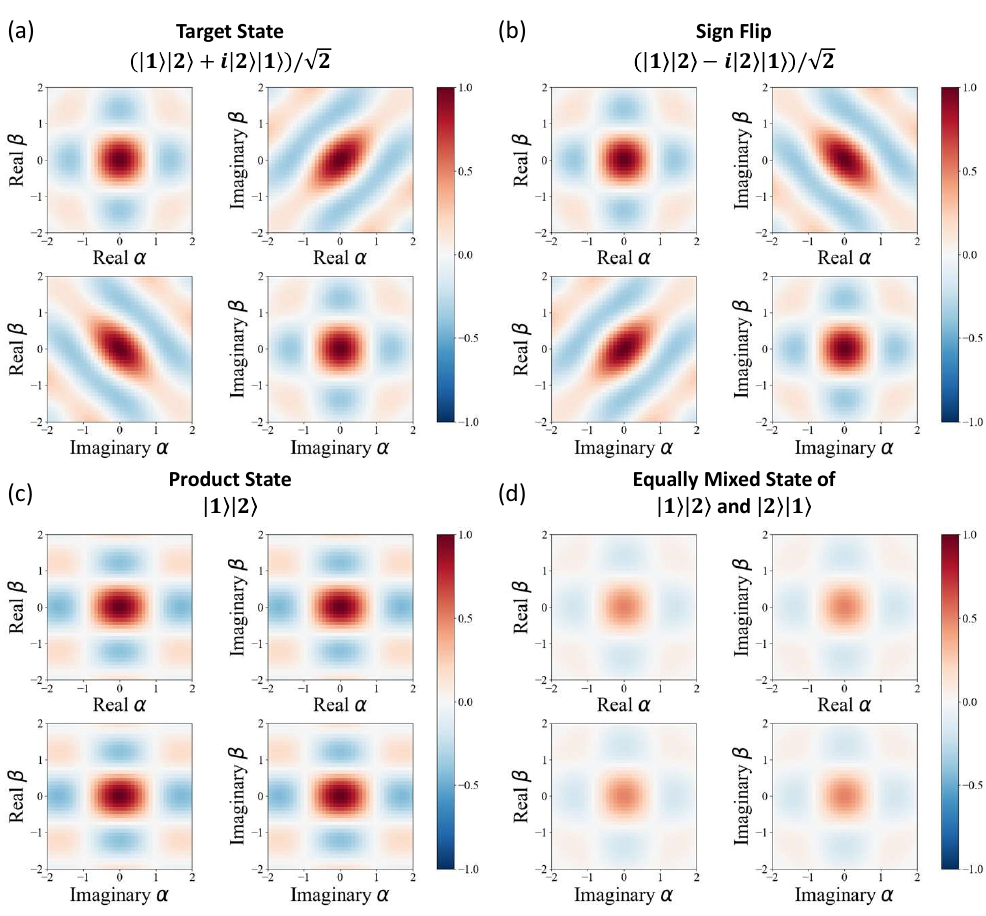}
    \caption{\label{panel_1_2}Comparison of the real part of the joint characteristic function for the desired target state \(\frac{1}{\sqrt{2}}(\ket{1}\ket{2}+i\ket{2}\ket{1})\) , representative phase errors, a product state and an incoherent mixture.}
\end{figure*}

We use Eq. (\ref{chara_equation}) to predict how phase errors and loss of coherence modify the characteristic-function slices measured in the main text. For the state \((\ket{0}_a(\ket{0}_b+\ket{1}_b) + ie^{i\Delta\theta} (\ket{0}_a+e^{i\Delta\phi}\ket{1}_a)\ket{0}_b)/\sqrt{\mathcal{N}}\), where $\Delta\theta$ is the possible phase error associated with the SWAP component, $\Delta\phi$ is the additional state-dependent phase error and $\mathcal{N}=4-2\sin\Delta\theta$ is the normalization factor, Eqs. (\ref{exp D}) and (\ref{chara_equation}) give

\begin{align}
\langle D_a(\alpha)D_b(\beta)\rangle&=e^{-\frac{\abs{\alpha}^2+\abs{\beta}^2}{2}}\qty(1-\frac{1}{\mathcal{N}}\abs{\Tilde{\kappa}}^2+\frac{4i\cos\qty(\frac{\Delta\theta}{2}+\frac{\pi}{4})}{\mathcal{N}}\Im(\Tilde{\kappa}))
\end{align}
where
\begin{align}
    \Tilde{\kappa}=\alpha e^{-i(\Delta\phi+\frac{\Delta\theta}{2}+\frac{\pi}{4})} +\beta e^{i(\frac{\Delta\theta}{2}+\frac{\pi}{4})}
    ~.
\end{align}
The imaginary part of this characteristic function is proportional to $\Im(\Tilde{\kappa})$. Consequently, phase errors rotate or reflect the antisymmetric features in the selected phase-space slices. For example, on the $\Re(\alpha)-\Re(\beta)$ plane, the antisymmetry axis satisfies $\alpha \sin(\Delta\phi+\frac{\Delta\theta}{2}+\frac{\pi}{4})=\beta \sin(\frac{\Delta\theta}{2}+\frac{\pi}{4})$. In the absence of phase errors, this reduces to \(\alpha=\beta\) as shown in Fig. \ref{panel_0_0+1}(a). A parity-induced phase error with \(\Delta\phi=\pi\) instead changes the antisymmetry axis to \(\alpha=-\beta\) as shown in Fig. \ref{panel_0_0+1}(c). The other phase-space planes provide complementary information and help distinguish \(\Delta\theta\) from \(\Delta\phi\). Other panels in Fig. \ref{panel_0_0+1}  show representative characteristic-function slices for different combinations of \(\Delta\theta\) and \(\Delta\phi\), and an incoherent mixture of the two components.

For the state \((|1,2\rangle+i e^{i\Delta\theta}|2,1\rangle)/\sqrt2\), Eq. (\ref{chara_equation}) reduces to the real-valued function, 
\begin{align}
    \langle D_a(\alpha)D_b(\beta)\rangle&=e^{-(|\alpha|^2+|\beta|^2)/2}\qty(F(|\alpha|^2,|\beta|^2)+\frac{1}{2}(2-|\alpha|^2)(2-|\beta|^2)\Im{\alpha^*\beta e^{i\Delta\theta}})
\end{align}
where $F(|\alpha|^2,|\beta|^2)$ is a rotationally symmetric function in each phase-space coordinate,
\begin{align*}
    F(|\alpha|^2,|\beta|^2)&=\frac{1}{2}(L_1(|\alpha|^2)L_2(|\beta|^2)+L_2(|\alpha|^2)L_1(|\beta|^2))
    ~.
\end{align*}
When \(\Delta\theta=0\), the mixed-quadrature slices of the characteristic function exhibit the symmetry patterns shown in Fig. \ref{panel_1_2}(a). A phase flip with \(\Delta\theta=\pi\) reverses the interference term and interchanges the corresponding symmetry axes, as shown in Fig. \ref{panel_1_2}(b). In contrast, the product state in Fig. \ref{panel_1_2}(c) and the incoherent mixture in Fig. \ref{panel_1_2}(d) lack coherence between \(|1,2\rangle\) and \(|2,1\rangle\), so the cross term vanishes and all four selected planes have identical symmetry.

\newpage
\subsection{Simulation Details}\label{simulation}
\begin{table}[h!]
    \centering
    \begin{tabular}{cccc}
    \toprule
      Modes   & Mode Frequency & Heating Rate & Coherence Time \\
      &(MHz)&$\qty(\text{s}^{-1})$&(ms)\\
      \midrule
      \midrule
        Radial OP & 1.16 & 0.0 &   10 \\
        Axial OP & 1.01 & 0.0 & 200  \\
        Axial IP & 0.59 & 17 & 100 \\
    \bottomrule
    \end{tabular}
    \caption{The parameters used for heating and dephasing simulations. The measured heating rate of the OP modes are 0.0 to the reported precision.}
    \label{tab:parameters}
\end{table}

The numerical simulations reported in the main text were performed using QuTiP~\cite{qutip5}. The simulated system consists of a two-level spin and three bosonic modes, each represented in the truncated Fock basis $\{\ket{0},\ldots,\ket{9}\}$. Convergence was checked by increasing the cutoff to $\ket{20}$, which produced no appreciable change in the simulated observables. The initial motional states were constructed using the \texttt{fock()} and \texttt{coherent()} functions. The density matrix was then evolved using the Lindblad master-equation solver \texttt{mesolve()} through the eSWAP sequence shown in Fig.~\ref{circuit diagram} of the main text. The beam splitter Hamiltonians follow Eqs.~(\ref{eq:H_BS}) and (\ref{eq:H_SBS}), and spin rotations were implemented using the corresponding $\hat{\sigma}_x$ and $\hat{\sigma}_y$ operators. A final spin rotation maps $\ket{+}$ to $\ket{g}$.

The pulse durations for $R_x(\pi)$, $XBS(\pi)$, and $U_{BS}(\pi)$ were set to the typical measured values of $10\,\mu\mathrm{s}$, $600\,\mu\mathrm{s}$, and $1.2\,\mathrm{ms}$, respectively. The $U_{BS}$ Hamiltonian also includes the residual spin-dependent Stark shift, with terms proportional to $\ket{g}\bra{g}$ and $\ket{e}\bra{e}$ separated by a measured detuning of $3\,\mathrm{Hz}$. The experimental spin-echo sequence was included by dividing each $U_{BS}$ pulse into two equal halves and applying an $R_y(\pi)$ rotation between the halves and another after the second half.

To simulate the characteristic-function readout, the full spin--motion state was evolved under a spin rotation $R_y(\phi)$ and the spin-dependent displacement Hamiltonians given in Eq.~(\ref{eq:H_D}) of the main text, with displacement amplitudes $\alpha/2$ and $\beta/2$ for modes $a$ and $b$, respectively. The three motional modes were then traced out using \texttt{ptrace()} to obtain the reduced spin density matrix $\rho_{\mathrm{s}}$. The simulated spin-excitation probability was calculated as $P_e=\operatorname{Tr}\!\left[\rho_{\mathrm{s}}\ket{e}\bra{e}\right]$, and used to calculate the simulated characteristic function from $1-2P_e$, as in the experiment.
The ideal characteristic function was calculated directly from the target two-mode density matrix $\rho_{\mathrm{target}}$ as
\(
\chi_{\mathrm{ideal}}(\alpha,\beta)
=
\operatorname{Tr}\!\left[
\rho_{\mathrm{target}}D_a(\alpha)D_b(\beta)
\right]
\).

Motional heating and dephasing were numerically simulated using independent collapse operators~\cite{sutherland2022one} for each mode $l\in\{a,b,c\}$,
\begin{equation}
L_{l,+}=\sqrt{\Gamma_l
}\,\hat{l}^\dagger,\qquad
L_{l,-}=\sqrt{\Gamma_l}\,\hat{l},\qquad
L_{l,\phi}=\sqrt{\frac{2}{\tau_l}}\,\hat{l}^\dagger \hat{l},
\end{equation}
where $\Gamma_l$ is the independently measured heating rate and $\tau_l$ is the measured exponential coherence-decay time, as shown in Table~\ref{tab:parameters}. With this convention, heating alone gives $d\langle n_l\rangle/dt=\Gamma_l$, while dephasing alone causes coherences between Fock states $\ket{0}_l$ and $\ket{1}_l$ to decay as $e^{-t/\tau_l}$. Additional simulations with each mechanism enabled separately were used to assess its contribution.

Our measured linear increase in mean phonon number supports the use of the heating model of classical electric-field noise. The measured exponential coherence decay motivates an effective Markovian dephasing model over the experimentally relevant timescale. 

\end{document}